\documentclass[reprint, amsmath,amssymb, aps, pra, longbibliography]{revtex4-1}

\usepackage{soul}
\usepackage{amsmath}   
\usepackage{amssymb}
\usepackage{graphicx}
\usepackage{dcolumn}
\usepackage{bm}
\usepackage{amsfonts}
\usepackage{subfigure}
\usepackage{array}
\usepackage{float}
\usepackage{color}
\usepackage[colorlinks=true,linkcolor=blue]{hyperref}%
\hypersetup{allcolors=blue}
\usepackage[normalem]{ulem}
\usepackage{xcolor}

\begin{document}

\title{Tractor Atom Interferometry}
\date{\today }

\author{A. Duspayev}
    \email{alisherd@umich.edu}
\affiliation{Department of Physics, University of Michigan, Ann Arbor, MI 48109, USA}
\author{G. Raithel}
\affiliation{Department of Physics, University of Michigan, Ann Arbor, MI 48109, USA}

\begin{abstract}
We propose a tractor atom interferometer (TAI) based on three-dimensional (3D) confinement and transport of split atomic wavefunction components in potential wells that follow programmed paths. The paths are programmed to split and recombine atomic wavefunctions at well-defined space-time points, guaranteeing closure of the interferometer.
Uninterrupted 3D confinement of the interfering wavefunction components in the tractor wells eliminates coherence loss due to wavepacket dispersion. Using Crank-Nicolson simulation of the time-dependent Schr\"odinger equation, we compute the quantum evolution of scalar and spinor wavefunctions in several TAI sample scenarios. The interferometric phases extracted from the wavefunctions allow us to quantify gravimeter sensitivity, for the TAI scenarios studied. We show that spinor-TAI supports matter-wave beam splitters that are more robust against non-adiabatic effects than their scalar-TAI counterparts. We confirm the validity of semiclassical path-integral phases taken along the programmed paths of the TAI. Aspects for future experimental realizations of TAI are discussed.     
\end{abstract}

\maketitle

\section{Introduction}

Since their first demonstrations~\cite{Carnal1991,  keith1991, riehle1991, Kasevich1991}, atom interferometers (AI)~\cite{croninreview, ai2020proceed} have become a powerful tool with a broad range of applications in tests of fundamental physics~\cite{Tarallo2014, Schlippert2014, Kovachy2015, Jaffe2017, Rosi2017}, precision measurements~\cite{Fixler74, hanneke2008, Parker191, morel2020, Xu745, barrett} and applied sciences~\cite{Menoret2018, Bongs2019, alzar}. 
A challenge in AI design is to achieve a high degree of sensitivity with respect to the measured quantity (e.g., an acceleration) while minimizing geometrical footprint of the apparatus and maximizing readout bandwidth to allow for practical applications. 
Previous work on AI includes free-space~\cite{asenbaum, Rosi2014, hamilton} and point-source~\cite{dickerson, hoth} AI experiments, as well as guided-wave AI experiments~\cite{wu2007,moan2020} and proposals~\cite{Davis2008, Zimmermann2019}. 
Free-space and point-source AIs typically employ atomic fountains or dropped atom clouds. 
The point-source method supports efficient readout and data reduction~\cite{chen2020}, enables high bandwidth, and affords efficiency in the partial-fringe regime.
Atomic fountains typically employed in free-space AI maximize interferometric time and, hence, increase sensitivity~\cite{asenbaum, Rosi2014, hamilton}, but require large experimental setups.
Guided-wave AIs offer compactness and are often used as Sagnac rotation sensors, but are susceptible to noise in the guiding potentials. 
In both free-space and guided-wave AI, wavepacket dynamics along unconfined degrees of freedom can cause wave-packet dispersion and failure to close, {\sl{i.e.}} the split wavepackets may fail to recombine in space-time.
Coherent recombination of split atomic wavefunctions upon their preparation and time-evolution remains challenging in recent AI studies~\cite{stickney2002, stickney, burke, stickney2008}.

Here we propose and analyze an AI method in which there are no unconfined degrees of freedom of the center-of-mass (COM) motion. The method relies on confining, splitting, transporting and re-combining atomic COM quantum states in three-dimensional (3D) quantum wells that move along user-programmed paths.
We refer to this approach as ``tractor atom interferometer" (TAI). Proper tractor path control ensures closure of the interferometer, and tight 3D confinement at all times during the AI loop suppresses  coherence loss due to wavepacket dispersion.

\begin{figure*}[!htb]
 \centering
  \includegraphics[width=\textwidth]{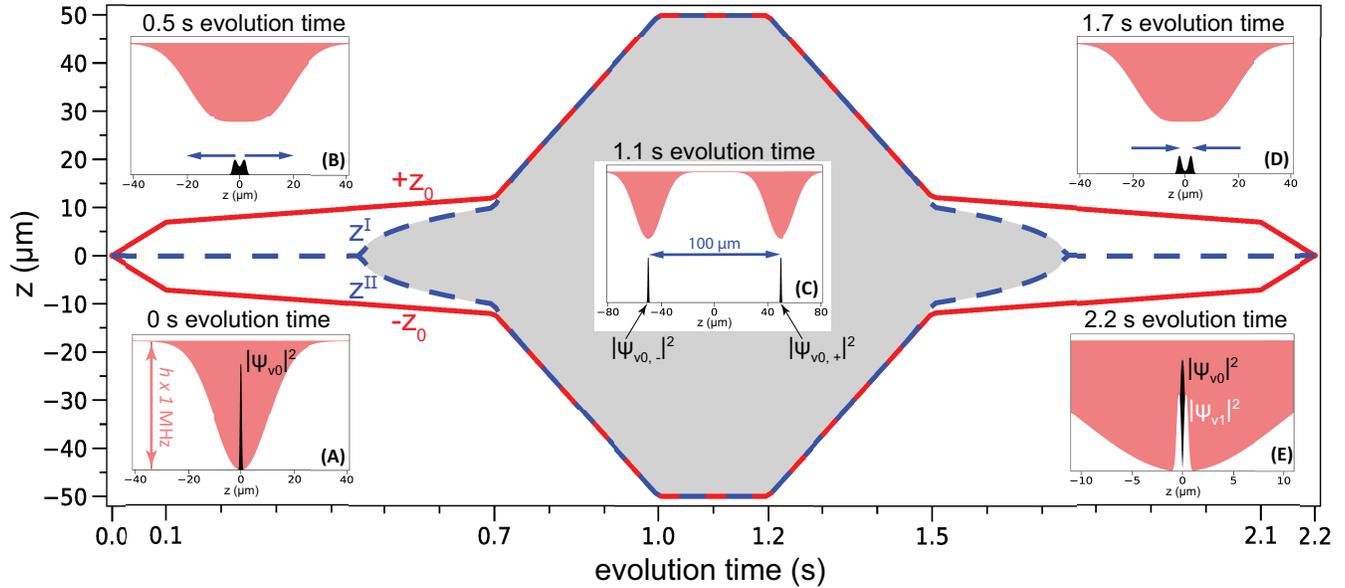}
  \caption{(Color online) Tractor control functions, $\pm z_0(t)$ (solid lines), and tractor paths, $z^{I/II}(t)$ (dashed lines), for a scalar TAI. The gray-shaded region shows the AI area.
  The insets (A)-(E) show how the trapping potential $U_1(z,t)$ and the scalar wavefunction evolve in time. Note the different $z$-ranges in the insets. The final state is a superposition of the ground ($\vert \psi_{\nu 0} \rangle $) and first excited ($\vert \psi_{\nu 1} \rangle $) COM states in the recombined well, as seen in inset (E), with the population ratio revealing the interferometric phase $\Delta \phi_Q$.}
  \label{scalar}
\end{figure*}

\section{Quantum model}

The quantum state of a single atom in the COM and spin product state space is
\begin{equation}
\vert \psi \rangle = \sum_{i=1}^{i_{\rm {max}}} \vert \psi_i (t) \rangle \otimes \vert i \rangle \quad ,
\label{eq:wfu1}
\end{equation}
with COM components $\vert \psi_i (t) \rangle$ in a number of spin states $i_{\rm {max}}$. For simplicity, we assume that all elements of the spin-state basis $\{ \vert i \rangle \}$ are position- and time-independent, and that the $x$- and $y$-degrees of freedom of the COM are frozen out. Denoting  $\psi_i(z,t) = \langle z \vert \psi_i (t) \rangle$, the time-dependent Schr\"odinger equation becomes
\begin{eqnarray}
{\rm {i}} \hbar \frac{\partial}{\partial t} \psi_i(z,t) & = & 
-\Big[ \frac{\hbar^2}{2m} \frac{\partial^2}{\partial z^2} + U_i(z,t) \Big]
 \psi_i(z,t) \nonumber \\
~ & ~ & + \sum_{j=1}^{i_{\rm {max}}} \frac{\hbar \Omega_{ij} (z,t)}{2} \psi_j(z,t) \quad ,
\label{eq:tdse}
\end{eqnarray}
with $i = 1, ..., i_{\rm{max}}$, particle mass $m$,  COM potentials $U_{i} (z, t)$ that may depend on spin, and couplings $\Omega_{ij}(z,t)$ between the spin states. 

In our examples below, we consider a scalar case, in which $i_{\rm {max}}=1$, and a spinor case with $i_{\rm {max}}=2$.
In the scalar case, the tractor-traps of TAI are all contained in a single potential $U_1 (z,t)$ for a scalar wavefunction $\psi_1 (z,t)$ (and there are no couplings $\Omega_{ij}$). In the spinor case, the spin space can be viewed as that of a spin-1/2 particle with spin states $\{ |\uparrow\rangle, |\downarrow\rangle \}$. The spin states could, for instance, represent two magnetic sublevels of the $F=1$ and $F=2$ hyperfine ground states of $^{87}$Rb. 
In spinor TAI, the spin states have distinct potentials, $U_{\uparrow} (z, t)$ and $U_{\downarrow} (z, t)$, with spin-specific potential wells, and the spinor wavefunction components are coupled via $\Omega_{\downarrow \uparrow}=\Omega^*_{\uparrow \downarrow}$.

We numerically solve Eq.~\ref{eq:tdse} using the Crank-Nicolson (CN) method~\cite{cnaref}. We use $^{87}$Rb atoms in wells $\sim20~\mu$m wide and $\sim h \times 1$~MHz deep. We employ a time step size of $\Delta t = $~10~ns, and a spatial grid step size of $\Delta z = $~10~nm. 
For the spinor simulations, we have generalized our CN algorithm to cover problems with $i_{\rm{max}} > 1$.

\section{Scalar TAI}
\label{sec:scalar}

In our scalar TAI implementation, the scalar potential $U_1(z,t)$ is the sum of two identically-shaped Gaussian potential wells that are both $h \times 500$~kHz deep and have a full-width-at-half-depth of 23.5~$\mu$m. The two Gaussians are centered at positions that are chosen to be symmetric in $z$ and are given by programmed tractor control functions $\pm z_0(t)$ (red solid lines in Fig.~\ref{scalar}). Initially, they are co-located at $z_0=0~\mu$m, forming a single trap that is $h\times1$~MHz-deep. A $^{87}$Rb atom is initialized in its COM ground state, $\vert \psi_{\nu 0} \rangle$,  of the $h\times1$~MHz-deep well (inset (A) in Fig.~\ref{scalar}). The function $z_0(t)$ is then gradually ramped up in order to split the single initial well into a pair of symmetric wells, causing the wavefunction to coherently split into two components (inset (B) in Fig.~\ref{scalar}). For $|z_0|\gtrsim 10~\mu$m, the split wells are about $h \times 500$~kHz deep. The minima in $U_1(z,t)$ follow the paths $z^I (t)$ and $z^{II} (t)$ shown by dashed blue lines in Fig.~\ref{scalar}, with $z^I (t) = -z^{II}(t)$. The paths $z^{I/II} (t)$ are found by solving 
$(\partial / \partial z) U_1 (z, t) = 0$. After the splitting, tunneling-induced coupling of the wavefunction components ceases, and the components adiabatically follow the separated paths of the potential minima. 
In our example, the wavefunction components stay separated for about 1~s, with the separation held constant at a maximum of $2 z_0 = 100~\mu$m for a duration of 0.2~s (inset (C) in Fig.~\ref{scalar}).
The wells and wavefunction components in them are recombined in a fashion that mirrors the separation (insets (D)  and (E)).
In order to provide a sufficient degree of adiabaticity, the ramp speed of the tractor control functions, $|\dot{z}_0(t)|$, is reduced near the times when the wavefunction splits and recombines.
The total duration of the cycle is 2.2~s, which is in line with the typical operation of modern AIs~\cite{Xu745}. 

In scalar TAI, the tractor paths, $z^{I/II}(t)$, differ significantly from the tractor control functions, $\pm z_0(t)$, during the well separation and recombination phases, while they are essentially the same when the minima are separated by more than the width of the wells (compare red-solid and blue-dashed lines in Fig.~\ref{scalar}). The AI area, visualized in Fig.~\ref{scalar} by a gray shading, is the area enclosed by the paths $z^{I}(t)$ and $z^{II}(t)$.

AI closure is guaranteed by virtue of proper tractor control. This is evident in the simulated wavefunction plots included in Fig.~\ref{scalar}. The quantum AI phase, $\Delta \phi_Q$, accumulates in the phases of the complex coefficients of the COM ground states in the split wells at positive and negative $z$, denoted $\vert \psi_{\nu 0,\pm} \rangle$ (inset (C) of Fig.~\ref{scalar}). 
Upon recombination of the pair of wells into a single well (insets (D) and (E) of Fig.~\ref{scalar}), the atomic state becomes mapped into a coherent superposition of the lowest and first-excited quantum states of the combined well, $\vert \psi \rangle = c_{\nu 0} \vert \psi_{\nu 0} \rangle + c_{\nu 1} \vert \psi_{\nu 1} \rangle$ (inset (E) of Fig.~\ref{scalar}). The observable probabilities, $|c_{\nu 0}|^2$ and $|c_{\nu 1}|^2$, yield the TAI quantum phase, $\Delta \phi_Q$, via the relation
\begin{equation}
    \frac{|c_{\nu 1}|^2}{|c_{\nu 0}|^2 + |c_{\nu 1}|^2} = \sin^2(\frac{\Delta \phi_Q}{2}),
    \label{scalarphi} \quad .
\end{equation}

\begin{figure}[t]
 \centering
  \includegraphics[width=0.49\textwidth]{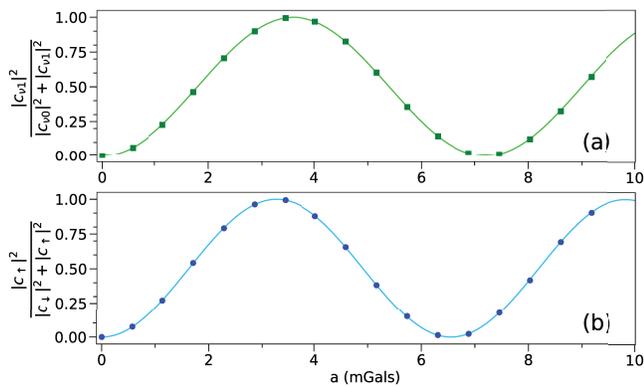}
  \caption{(Color online) Simulated measurement of $\sin^2(\frac{\Delta \phi_Q}{2})$ vs a small background acceleration, $a$, using TAI with scalar (a) and spinor (b) wavefunctions. 
  Squares and dots in (a) and (b) are from wavefunction simulation data, while solid lines are semiclassical path-integral data. No fitting is performed.}
  \label{results}
\end{figure}

Due to uninterrupted 3D confinement, TAI eliminates free-particle wave-packet dispersion. There is, however, a possibility of non-adiabatic transitions into excited COM states during the wavefunction splitting and recombination, which would reduce the interferometer's contrast and introduce spurious signals. To show that under conditions such as in Fig.~\ref{scalar} there is no significant coherence loss due to non-perfect closure or non-adiabatic effects, we have performed wavefunction simulations for 17 values of a small acceleration, $a$, along the $z$-direction. The acceleration adds a potential $U_g = m\, a \,z$ to the TAI potential $U_1(z,t)$. We vary $a$ over a range from 0 to 10~mGal ($10^{-4}$~m/s$^2$). From the wavefunction simulations, we determine $\sin^2(\Delta \phi_Q /2)$ according to Eq.~\ref{scalarphi} as a function of $a$, and plot the results in Fig.~\ref{results}~(a) (symbols). The visibility of the expected sinusoidal dependence reaches near-unity, providing evidence for near-perfect closure and absence of coherence loss due to non-adiabatic COM excitations. The acceleration sensitivity is discussed in Sec.~\ref{sec:sensnonad}.

\section{Spinor TAI}
\label{sec:spinor}

The scalar scheme in Sec.~\ref{sec:scalar} serves well to describe the TAI concept. At the splitting, the initial COM state is supposed to evolve into the even-parity superposition of the ground states in the split wells, $(\vert \psi_{\nu 0, +} \rangle +
\vert \psi_{\nu 0, -} \rangle)/\sqrt{2}$, without populating the odd-parity superposition and other excited COM states. However, under conditions that are less ideal than in  Sec.~\ref{sec:scalar}, scalar TAI is prone to  non-adiabatic excitation of unwanted COM states at the times when the wells split and recombine.
The splitting and, similarly, the recombination are fragile because the potential is very soft at the splitting and recombination times, and non-adiabatic mixing can easily occur (see Sec.~\ref{sec:sensnonad}).

\begin{figure*}[t]
 \centering
  \includegraphics[width=\textwidth]{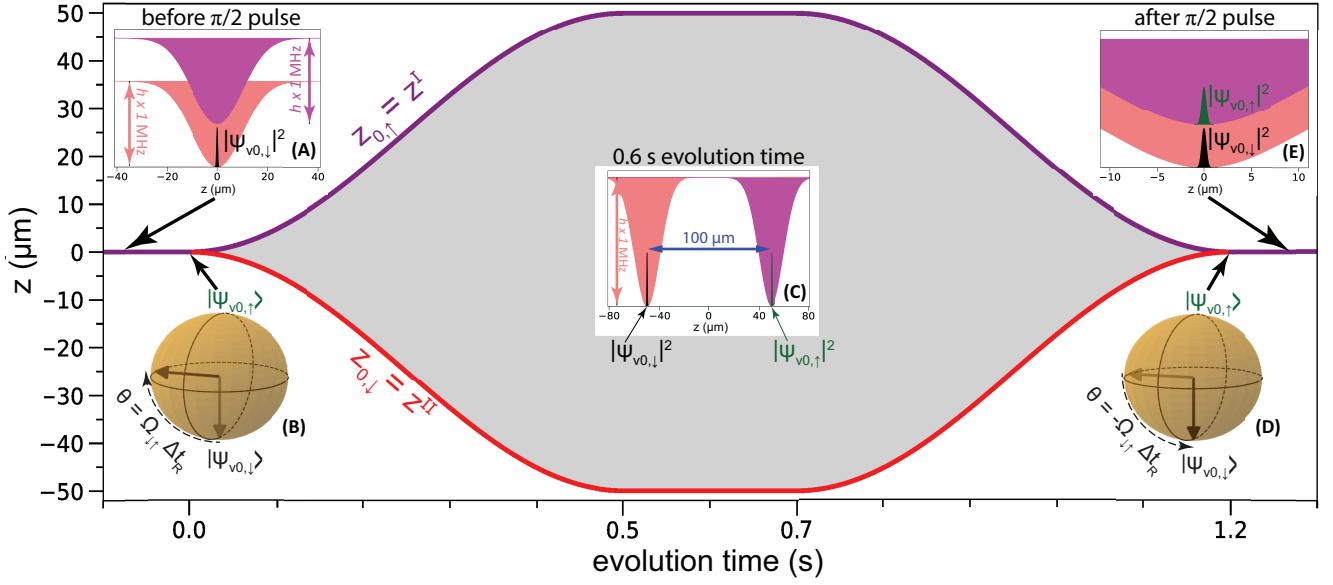}
  \caption{(Color online) 
 Example for spinor TAI. The tractor control functions, $z_{0,\downarrow}(t)$ and  $z_{0,\uparrow}(t)$, are identical with the respective paths, $z^{II}(t)$ and $z^{I}(t)$. The interferometric area is shaded in gray. The insets
 (A), (C), (E) show spin-dependent tractor potentials and spinor wavefunctions at different instants of the evolution time. Note the different $z-$ranges in these insets. 
 The Bloch spheres in insets (B) and (D) visualize the $\pm \pi/2$-splitting and recombination pulses of spinor TAI. The final-state populations in the spin states reveal the interferometric phase $\Delta \phi_Q$.}
  \label{spinor}
\end{figure*}

The fragility of scalar TAI is avoided in our second, improved method that operates on a two-component spinor system ($i_{\rm{max}} = 2$ in Eqs.~\ref{eq:wfu1} and~\ref{eq:tdse}) with a pair of spin-dependent potentials. The atomic wavefunction is initially prepared in the COM ground state $\vert \psi_{\nu 0, \downarrow} \rangle$ of a spin-down tractor potential (inset (A) in Fig.~\ref{spinor}). With initially overlapping and identical spin-down and -up tractor potentials, $U_{\downarrow}(z,t=0)$ and $U_{\uparrow}(z,t=0)$, a short $\pi/2$ coupling pulse with Rabi frequency $\Omega_{\downarrow \uparrow}$ (see Eq.~\ref{eq:tdse}) prepares a coherent superposition $(\vert \psi_{\nu 0, \downarrow} \rangle + \vert \psi_{ \nu 0, \uparrow} \rangle)/\sqrt{2}$ of the COM ground states of the potentials $U_{\downarrow}(z,t=0)$ and $U_{\uparrow}(z,t=0)$.
The $\pi/2$-pulse can be realized by a microwave or a momentum-transfer-free optical Raman transition. 
In the simulated case, the Rabi frequency $\Omega_{\downarrow \uparrow}$ is position-independent, has a fixed magnitude $\Omega_{\downarrow \uparrow}(t) = 2 \pi \times 178$~kHz, and is on for a coupling-pulse duration of $\Delta t_R = 1.4~\mu$s, which is short on the time scale of the interferometer.
The resultant TAI splitting is depicted on the Bloch sphere in the inset (B) of Fig.~\ref{spinor}. 
After the $\pi/2$-pulse, the spin-dependent potentials and the spinor wavefunction components in them are translated following symmetric tractor control functions $z_{0, \uparrow}(t) = z_{0}(t) = - z_{0,\downarrow}(t)$.
After a holding time of 0.2~s at a maximal separation of 100~$\mu$m (inset (C) in Fig.~\ref{spinor}), the splitting is reversed and the wavefunction components become overlapped again. 
Closure occurs via an exit $-\pi/2$ pulse (inset (D) in Fig.~\ref{spinor}). In the case of spinor TAI, the tractor paths and control functions are identical,  $z^{I}(t) = z_{0, \downarrow}(t)$ and $z^{II}(t) = z_{0, \uparrow}(t)$. The AI area is visualized by the gray shaded region in Fig.~\ref{spinor}.

In the absence of non-adiabatic transitions into excited COM states in the spin-dependent potentials, the final state is of the form 
$\vert \psi \rangle = c_{\downarrow} \vert \psi_{ \nu 0, \downarrow} \rangle + c_{\uparrow} \vert \psi_{\nu 0, \uparrow} \rangle$.  The AI phase $\Delta \phi_Q$ is encoded in the final populations in the two spin states (inset (E) in Fig.~\ref{spinor}) and follows
\begin{equation}
    \frac{|c_{\uparrow}|^2}{|c_{\downarrow}|^2 + |c_{\uparrow}|^2} = \sin^2(\frac{\Delta \phi_Q}{2}),
    \label{spinorphi} \quad .
\end{equation}
In experimental implementations,
the $|c_{\downarrow}|^2$ and $|c_{\uparrow}|^2$ can be measured, for instance, via state-dependent fluorescence to yield $\sin^2(\frac{\Delta \phi_Q}{2})$.

Similar to the scalar case in Sec.~\ref{sec:scalar}, we have performed wavefunction simulations for a set of accelerations $a$ along the $z$-direction, which add identical gravitational potentials $U_g = m\, a \,z$ to both spin-dependent potentials. From the simulated spinor wavefunctions we extract $c_{\uparrow}$ and $c_{\downarrow}$, compute $\sin^2(\Delta \phi_Q /2)$ according to Eq.~\ref{spinorphi}, and plot the results in Fig.~\ref{results}~(b) (symbols). The results again provide evidence for near-perfect closure and absence of coherence loss due to non-adiabatic COM transitions. 

\section{Comparison of quantum and semiclassical phases} 
\label{sec:semi}

Using the path integral formalism, the semiclassical phase of an AI loop, $\Delta\phi_S$, in 1D is~\cite{croninreview}:
\begin{equation}
    \Delta\phi_S=\frac{\int_{t_a}^{t_b} [\mathcal{L}^{II}(z, \dot{z}, t) - \mathcal{L}^{I}(z, \dot{z}, t)]~dt}{\hbar},
    \label{classicaction}
\end{equation}
\noindent where $\Delta\phi$ is in rads, $\mathcal{L}^{II/I}$ are the Lagrange functions on the paths $z^{II/I}(t)$ of the centroids of the split atomic wavefunction components, and $t_a$ and $t_b$ are the splitting and recombination times. 

A key feature that distinguishes TAI from other AIs is that the paths $z^{II/I}(t)$ are pre-determined by the system controls (and therefore do not have to be computed prior to using Eq.~\ref{classicaction}). 
Simultaneous arrival of the split wavefunction components at the recombination point is achieved by proper programming of the tractor paths. 

The guaranteed closure of TAI in space-time is related to the fact that the number of generalized Lagrangian coordinates in TAI is zero. Other AIs typically have at least one generalized coordinate, along which the classical motion is unconstrained and along which quantum wavepackets may disperse. The AI can then, in principle, fail to achieve closure due to a difference in classical propagation times along the AI paths between splitting and recombination.
A propagation time difference can be caused by uncontrollable conditions, such as an erratic background acceleration. In TAI, closure is guaranteed  by virtue of uninterrupted 3D control of the interferometric paths and suitable tractor programming.

In our examples we have considered tractor paths in which the kinetic energy terms in $\mathcal{L}^{II/I}$ are equal, {\sl{i.e.}} $\dot{z}^{II}(t)= -\dot{z}^{I}(t)$, and we have added a gravitational potential $U_g = m a z$. In that case, Eq.~\ref{classicaction} simplifies to
\begin{equation}
    \Delta\phi_S=\frac{\int_{t_a}^{t_b} [U_g(z^{I}(t))-U_g(z^{II}(t))]~dt}{\hbar}=\frac{m \, a \, C}{\hbar}.
    \label{phase}
\end{equation}
\noindent with a parameter
\begin{equation}
    C = \int_{t_a}^{t_b} [z^{I}(t)-z^{II}(t)]~dt
    \label{Cdef} \quad ,
\end{equation}
\noindent that only depends on the programmed tractor paths $z^I(t)$ and $z^{II}(t)$. Note there is no atom dynamics to be solved for. The $z^{I}(t)$ and $z^{II}(t)$ are either identical with the tractor control functions $z_{0,*}(t)$ themselves (spinor case), or they are found by solving an equation of the type $(\partial / \partial z) U_1 (z, t) = 0$ (scalar case).

We compare the semiclassical phases $\Delta \phi_S$ with the quantum phases $\Delta \phi_Q$ over a range of accelerations, $a$.
The $\Delta \phi_S (a)$ that follow from Eqs.~\ref{classicaction}-\ref{Cdef} 
after utilization of the appropriate tractor paths $z^{I}(t)$ and $z^{II}(t)$ are shown in Figs.~\ref{results}~(a) and (b) as solid lines.
We find in both cases that $\Delta \phi_S = \Delta \phi_Q$, with minor discrepancies that are not visible in the figure. Quantum and semiclassical phases are both offset-free, {\sl{i.e.}} in our case of symmetric controls and for $a=0$ it is $\Delta \phi_S = \Delta \phi_Q = 0$. This indicates that the exact quantum dynamics does not add an offset splitting and recombination phase.

\section{Sensitivity, computation accuracy and non-adiabatic effects}
\label{sec:sensnonad}

The scalar and spinor implementations simulated in Secs.~\ref{sec:scalar} and Sec.~\ref{sec:spinor} exhibit similar sensitivities to the acceleration $a$. The sensitivities are not the same because the cases happen to have slightly different AI areas (shaded regions in Figs.~\ref{scalar} and~\ref{spinor}).
Assuming a phase resolution of $2\pi/100$, the acceleration sensitivity of the sequences in Figs.~\ref{scalar} and~\ref{spinor} is on the order of several hundreds of microGals ($\sim 10^{-7}g$), which is about a factor of 100 short of the level of modern gravimeters~\cite{Menoret2018}. TAI could reach that level by a ten-fold increase of the interferometer time, $T_I = t_b-t_a$, ([1~s, 1.2~s] in Fig.~\ref{scalar} and [0.5~s, 0.7~s] in Fig.~\ref{spinor}) and a ten-fold increase of the spatial separation between the tractor potential wells.

We have checked that a reduction of the grid spacing $\Delta z$ in the simulation does not noticeably affect the accuracy of the results, whereas a reduction of the time step size $\Delta t$ does improve the agreement of $\Delta \phi_S$ with $\Delta \phi_Q$. Therefore we attribute the minor differences between quantum and semiclassical phases (too small to be seen in Figs.~\ref{results}~(a) and (b)) mostly to the finite step size, $\Delta t=10$~ns, in the CN simulation.
The step-size parameters chosen in our work reflect a trade-off between accuracy of $\Delta \phi_Q$ and simulation time needed.

It remains to be seen in the future whether quantum and Lagrangian phases may exhibit offsets of a physical nature. Such offsets may occur, for instance, if the tractor control functions $z_{0,*}(t)$ were asymmetric, if the tractor-trap paths $z^{I/II}(t)$ had accelerations large enough to cause wavepacket drag in the 3D tractor wells, or if trajectory bundles of the semi-classical AI  traversed through caustics or focal spots that lead to quantum phase shifts.

We have noted that scalar TAI generally is more susceptible to non-adiabatic COM excitations in the splitters and recombiners than spinor TAI, necessitating longer splitter- and recombiner durations with reduced slopes $|\dot{z_0}|$ near the critical time points when the single well splits into two and vice versa (see Fig.~\ref{scalar}). 
This entails a longer overall AI sequence, a reduced reading bandwidth, and additional susceptibilities to noise (such as vibrations) during the splitting and recombination. 
These shortcomings are naturally avoided in spinor TAI, where the tractor potentials do not soften at the splitting and recombination times.

\begin{figure}[t]
 \centering
  \includegraphics[width=0.49\textwidth]{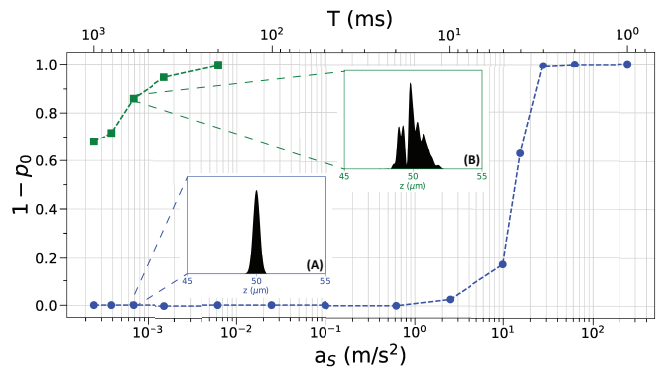}
  \caption{(Color online) Non-adiabaticity vs splitting duration $T$ (top axis) and peak tractor acceleration $a_S$ (bottom axis) for a tractor control function given in the text, for scalar (green squares) and spinor TAI (blue circles). The dashed lines connecting the respective data points are to guide the eye. It is seen that spinor TAI (blue circles) allows about a factor of 100 faster splitting than scalar TAI. The insets show representative wavefunction densities after the splitting for a tractor acceleration in which spinor-TAI performs very well while scalar TAI  fails.}
  \label{results2}
\end{figure}

To quantify the non-adiabaticity in both TAI cases, we have run a series of simulations of splitting sequences with smooth tractor control functions  $z_0(t) = 50~\mu$m$ \times \sin^2{[\pi t/(2T)]}$ (as in Fig.~\ref{spinor}) for a range of splitter durations $T$ and acceleration $a=0$.
The non-adiabaticy is given by $1-p_0(T)$, where $p_0$ is the COM ground-state probability after the splitting.
In Fig.~\ref{results2} we plot $1-p_0(T)$ vs $T$ and the peak value, $a_S=25 \mu$m$\,(\pi/T)^2$, of the splitter acceleration $|\ddot{z}_0|$ for both TAI cases. 
The wavefunction densities in the inset visualize the contrast between adiabatic (inset (A); no COM excitation) and non-adiabatic splitting (inset (B); substantial COM excitation).
The results underscore that for scalar TAI it is crucial to reduce the slope $|\dot{z}_0|$ at the times when the wells split and recombine.
For the control-function type used in Fig.~\ref{results2}, spinor TAI allows for rapid splitting, $T\sim 10$~ms, while scalar TAI requires splitter times $T\sim 1$~s.  

For Sagnac rotation interferometry~\cite{wu2007,moan2020,dickerson,chen2020}, the tractor paths can be programmed to circumscribe a non-zero geometric area $A$, and the paths can be traversed $N$ times between splitting and recombination. For a sensitivity estimate for the angular rotation rate $\Omega$, we assume TAI loop parameters of $A=1$~cm$^2$ and $N=300$, which seems feasible. For rubidium it then is $\Delta \phi / \Omega \sim m A / \hbar \approx 4 \times 10^7$ rad/(rad/s). Assuming a phase resolution $\Delta \phi = 2\pi/100$, the rotation sensitivity would be $\sim$1~nrad/s. 

\section{Discussion} 

In this Section, we consider aspects regarding experimental implementations of TAI. Useful methods will likely include elements of pioneering demonstrations of AI with Bose-Einstein condensates~\cite{shin}, of proposals for AI in magnetic microtraps~\cite{hansel,hinds,hanselpra}, and of emerging techniques of moving (arrays of) ultracold atoms on 
arbitrary trajectories~\cite{Endres1024, Barredo1021, kim2016, ohl, schymik}. To reduce quantum projection noise, the utilized method must allow for parallel operation of many identical loops in a small overall volume~\cite{kovachypra2010}. Compact, 3D-confining and parallelizable platforms that allow dynamic tractor control include optical lattices~\cite{Mandel2003, Kuhr2001, Schmid2006, Kumar2018} and optical tweezers~\cite{Endres1024, Barredo1021, kim2016,  schymik} that may use an array of micro-lenses~\cite{ohl}. 

TAI can fail if the acceleration to be measured is too large, causing splitter asymmetry and exaggerated non-adiabatic effects. In~\cite{Kovachy2015}, $a$ needed to be $\lesssim 10^3$~Gals $\approx $1~$g$.
With the quantum confinement in 3D traps, afforded by TAI, the tractor traps can be designed steep enough to avoid such effects, which can be an advantage in high-$g$ scenarios.
As seen in Fig.~\ref{results2}, where traps of a depth of $h \times$500~kHz and well sizes of 20~$\mu$m are used, spinor-TAI is especially promising in that regard. Deeper and tighter traps in optical lattices with well sizes $<1~\mu$m  are expected to afford closure and robustness under high-$g$ conditions.  

As in guided-wave AI, in TAI it is important to avoid differential fluctuations in depth and position of the tractor potentials. In experimental implementations, this will present a serious challenge. For instance, the sensitivity to differential noise in the potential depths, $\delta \Delta V = \delta (V^I - V^{II})$, with $V^I$ and $V^{II}$ denoting the depths of the individual tractor wells, scales with the interferometer time $T_I = t_b - t_a$, which is $\sim 1$~s in our examples. The requirement  $\delta \Delta V T_I < h$ then leads to a requirement of $\delta \Delta V < h \times 1$~Hz. For wells that are $h \times 100~$kHz deep, the allowed variation in differential tractor depth therefore is in the range of $10^{-5}$ of the well depth. 
This estimation becomes considerably more favorable for tractor wells that are approximations of deep square wells with an absolute trapping potential near zero inside the wells. One may envision, for instance, traps formed by blue-detuned optical lattices or box potentials, in which the atoms are trapped near locations of minimal light intensity.

Along similar lines, in sensing applications where common-mode fluctuations do not affect $\Delta \phi$~\cite{chiow2009}, TAI could be implemented with tractor controls that act symmetrically in both TAI paths. The symmetry requirement extends to several types of noise, including noise in the differential trap depths of the tractor wells, $\delta \Delta V$, in the differential potential energy of the tractor paths, $\delta \Delta U = m {\bf{a}} \cdot \delta ({\bf{r}}^I - {\bf{r}}^{II})$, in constant-acceleration background potentials, and in the differential kinetic energy along the tractor paths. The latter aspect translates into effects of differential mirror vibrations, phases of optical-lattice laser beams, etc. Future experimental work as well as detailed simulations that include various types of noise will help in exploring the opportunities afforded by TAI as well as in establishing its practical limitations. 

In summary, our theoretical study and the discussion  regarding experimental feasibility show that TAI 
is a promising concept for future experimental demonstration. TAI offers a set of values that we believe is hard to match with free-space and partially-confining atom interferometers, including guaranteed closure in space-time, even under rough conditions with large and variable background accelerations and rotations, suppression of wavepacket dispersion due to uninterrupted 3D confinement, and use of user-programmable loops with arbitrary hold times and flexible geometries for a variety of sensing applications.

\maketitle
\section*{ACKNOWLEDGMENTS}
The work was supported by NSF under Grant No. PHYS-1806809 and NASA under Grant No. NNH13ZTT002N NRA. We thank Lu Ma and Vladimir Malinovsky for useful discussions.

\bibliography{references}

\begin{thebibliography}{50}%
\makeatletter
\providecommand \@ifxundefined [1]{%
 \@ifx{#1\undefined}
}%
\providecommand \@ifnum [1]{%
 \ifnum #1\expandafter \@firstoftwo
 \else \expandafter \@secondoftwo
 \fi
}%
\providecommand \@ifx [1]{%
 \ifx #1\expandafter \@firstoftwo
 \else \expandafter \@secondoftwo
 \fi
}%
\providecommand \natexlab [1]{#1}%
\providecommand \enquote  [1]{``#1''}%
\providecommand \bibnamefont  [1]{#1}%
\providecommand \bibfnamefont [1]{#1}%
\providecommand \citenamefont [1]{#1}%
\providecommand \href@noop [0]{\@secondoftwo}%
\providecommand \href [0]{\begingroup \@sanitize@url \@href}%
\providecommand \@href[1]{\@@startlink{#1}\@@href}%
\providecommand \@@href[1]{\endgroup#1\@@endlink}%
\providecommand \@sanitize@url [0]{\catcode `\\12\catcode `\$12\catcode
  `\&12\catcode `\#12\catcode `\^12\catcode `\_12\catcode `\%12\relax}%
\providecommand \@@startlink[1]{}%
\providecommand \@@endlink[0]{}%
\providecommand \url  [0]{\begingroup\@sanitize@url \@url }%
\providecommand \@url [1]{\endgroup\@href {#1}{\urlprefix }}%
\providecommand \urlprefix  [0]{URL }%
\providecommand \Eprint [0]{\href }%
\providecommand \doibase [0]{http://dx.doi.org/}%
\providecommand \selectlanguage [0]{\@gobble}%
\providecommand \bibinfo  [0]{\@secondoftwo}%
\providecommand \bibfield  [0]{\@secondoftwo}%
\providecommand \translation [1]{[#1]}%
\providecommand \BibitemOpen [0]{}%
\providecommand \bibitemStop [0]{}%
\providecommand \bibitemNoStop [0]{.\EOS\space}%
\providecommand \EOS [0]{\spacefactor3000\relax}%
\providecommand \BibitemShut  [1]{\csname bibitem#1\endcsname}%
\let\auto@bib@innerbib\@empty
%</preamble>
\bibitem [{\citenamefont {Carnal}\ and\ \citenamefont
  {Mlynek}(1991)}]{Carnal1991}%
  \BibitemOpen
  \bibfield  {author} {\bibinfo {author} {\bibfnamefont {O.}~\bibnamefont
  {Carnal}}\ and\ \bibinfo {author} {\bibfnamefont {J.}~\bibnamefont
  {Mlynek}},\ }\bibfield  {title} {\enquote {\bibinfo {title} {Young's
  double-slit experiment with atoms: A simple atom interferometer},}\ }\href
  {\doibase 10.1103/PhysRevLett.66.2689} {\bibfield  {journal} {\bibinfo
  {journal} {Phys. Rev. Lett.}\ }\textbf {\bibinfo {volume} {66}},\ \bibinfo
  {pages} {2689--2692} (\bibinfo {year} {1991})}\BibitemShut {NoStop}%
\bibitem [{\citenamefont {Keith}\ \emph {et~al.}(1991)\citenamefont {Keith},
  \citenamefont {Ekstrom}, \citenamefont {Turchette},\ and\ \citenamefont
  {Pritchard}}]{keith1991}%
  \BibitemOpen
  \bibfield  {author} {\bibinfo {author} {\bibfnamefont {D.~W.}\ \bibnamefont
  {Keith}}, \bibinfo {author} {\bibfnamefont {C.~R.}\ \bibnamefont {Ekstrom}},
  \bibinfo {author} {\bibfnamefont {Q.~A.}\ \bibnamefont {Turchette}}, \ and\
  \bibinfo {author} {\bibfnamefont {D.~E.}\ \bibnamefont {Pritchard}},\
  }\bibfield  {title} {\enquote {\bibinfo {title} {An interferometer for
  atoms},}\ }\href {\doibase 10.1103/PhysRevLett.66.2693} {\bibfield  {journal}
  {\bibinfo  {journal} {Phys. Rev. Lett.}\ }\textbf {\bibinfo {volume} {66}},\
  \bibinfo {pages} {2693--2696} (\bibinfo {year} {1991})}\BibitemShut {NoStop}%
\bibitem [{\citenamefont {Riehle}\ \emph {et~al.}(1991)\citenamefont {Riehle},
  \citenamefont {Kisters}, \citenamefont {Witte}, \citenamefont {Helmcke},\
  and\ \citenamefont {Bord\'e}}]{riehle1991}%
  \BibitemOpen
  \bibfield  {author} {\bibinfo {author} {\bibfnamefont {F.}~\bibnamefont
  {Riehle}}, \bibinfo {author} {\bibfnamefont {Th.}\ \bibnamefont {Kisters}},
  \bibinfo {author} {\bibfnamefont {A.}~\bibnamefont {Witte}}, \bibinfo
  {author} {\bibfnamefont {J.}~\bibnamefont {Helmcke}}, \ and\ \bibinfo
  {author} {\bibfnamefont {Ch.~J.}\ \bibnamefont {Bord\'e}},\ }\bibfield
  {title} {\enquote {\bibinfo {title} {Optical {R}amsey spectroscopy in a
  rotating frame: Sagnac effect in a matter-wave interferometer},}\ }\href
  {\doibase 10.1103/PhysRevLett.67.177} {\bibfield  {journal} {\bibinfo
  {journal} {Phys. Rev. Lett.}\ }\textbf {\bibinfo {volume} {67}},\ \bibinfo
  {pages} {177--180} (\bibinfo {year} {1991})}\BibitemShut {NoStop}%
\bibitem [{\citenamefont {Kasevich}\ and\ \citenamefont
  {Chu}(1991)}]{Kasevich1991}%
  \BibitemOpen
  \bibfield  {author} {\bibinfo {author} {\bibfnamefont {M.}~\bibnamefont
  {Kasevich}}\ and\ \bibinfo {author} {\bibfnamefont {S.}~\bibnamefont {Chu}},\
  }\bibfield  {title} {\enquote {\bibinfo {title} {Atomic interferometry using
  stimulated {R}aman transitions},}\ }\href {\doibase
  10.1103/PhysRevLett.67.181} {\bibfield  {journal} {\bibinfo  {journal} {Phys.
  Rev. Lett.}\ }\textbf {\bibinfo {volume} {67}},\ \bibinfo {pages} {181--184}
  (\bibinfo {year} {1991})}\BibitemShut {NoStop}%
\bibitem [{\citenamefont {Cronin}\ \emph {et~al.}(2009)\citenamefont {Cronin},
  \citenamefont {Schmiedmayer},\ and\ \citenamefont
  {Pritchard}}]{croninreview}%
  \BibitemOpen
  \bibfield  {author} {\bibinfo {author} {\bibfnamefont {A.~D.}\ \bibnamefont
  {Cronin}}, \bibinfo {author} {\bibfnamefont {J.}~\bibnamefont
  {Schmiedmayer}}, \ and\ \bibinfo {author} {\bibfnamefont {D.~E.}\
  \bibnamefont {Pritchard}},\ }\bibfield  {title} {\enquote {\bibinfo {title}
  {Optics and interferometry with atoms and molecules},}\ }\href {\doibase
  10.1103/RevModPhys.81.1051} {\bibfield  {journal} {\bibinfo  {journal} {Rev.
  Mod. Phys.}\ }\textbf {\bibinfo {volume} {81}},\ \bibinfo {pages}
  {1051--1129} (\bibinfo {year} {2009})}\BibitemShut {NoStop}%
\bibitem [{\citenamefont {Abend}\ \emph {et~al.}(2020)\citenamefont {Abend},
  \citenamefont {Gersemann}, \citenamefont {Schubert}, \citenamefont {Rasel},
  \citenamefont {Zimmermann}, \citenamefont {Efremov}, \citenamefont {Roura},
  \citenamefont {Narducci},\ and\ \citenamefont {Schleich}}]{ai2020proceed}%
  \BibitemOpen
  \bibfield  {author} {\bibinfo {author} {\bibfnamefont {S.}~\bibnamefont
  {Abend}}, \bibinfo {author} {\bibfnamefont {M.}~\bibnamefont {Gersemann}},
  \bibinfo {author} {\bibfnamefont {D.}~\bibnamefont {Schubert}, \bibfnamefont
  {C.~Schlippert}}, \bibinfo {author} {\bibfnamefont {E.~M.}\ \bibnamefont
  {Rasel}}, \bibinfo {author} {\bibfnamefont {M.}~\bibnamefont {Zimmermann}},
  \bibinfo {author} {\bibfnamefont {M.~A.}\ \bibnamefont {Efremov}}, \bibinfo
  {author} {\bibfnamefont {A.}~\bibnamefont {Roura}}, \bibinfo {author}
  {\bibfnamefont {F.~A.}\ \bibnamefont {Narducci}}, \ and\ \bibinfo {author}
  {\bibfnamefont {W.~P.}\ \bibnamefont {Schleich}},\ }\bibfield  {title}
  {\enquote {\bibinfo {title} {Atom interferometry and its applications},}\
  }in\ \href@noop {} {\emph {\bibinfo {booktitle} {Proceedings of the
  International School of Physics “Enrico Fermi”}}},\ \bibinfo {editor}
  {edited by\ \bibinfo {editor} {\bibfnamefont {W.~P.}\ \bibnamefont
  {Schleich}}, \bibinfo {editor} {\bibfnamefont {E.~M.}\ \bibnamefont {Rasel}},
  \ and\ \bibinfo {editor} {\bibfnamefont {S.}~\bibnamefont {W{\"o}lk}}}\
  (\bibinfo  {publisher} {IOS Press},\ \bibinfo {address} {Amsterdam},\
  \bibinfo {year} {2020})\BibitemShut {NoStop}%
\bibitem [{\citenamefont {Tarallo}\ \emph {et~al.}(2014)\citenamefont
  {Tarallo}, \citenamefont {Mazzoni}, \citenamefont {Poli}, \citenamefont
  {Sutyrin}, \citenamefont {Zhang},\ and\ \citenamefont {Tino}}]{Tarallo2014}%
  \BibitemOpen
  \bibfield  {author} {\bibinfo {author} {\bibfnamefont {M.~G.}\ \bibnamefont
  {Tarallo}}, \bibinfo {author} {\bibfnamefont {T.}~\bibnamefont {Mazzoni}},
  \bibinfo {author} {\bibfnamefont {N.}~\bibnamefont {Poli}}, \bibinfo {author}
  {\bibfnamefont {D.~V.}\ \bibnamefont {Sutyrin}}, \bibinfo {author}
  {\bibfnamefont {X.}~\bibnamefont {Zhang}}, \ and\ \bibinfo {author}
  {\bibfnamefont {G.~M.}\ \bibnamefont {Tino}},\ }\bibfield  {title} {\enquote
  {\bibinfo {title} {Test of {E}instein equivalence principle for 0-spin and
  half-integer-spin atoms: search for spin-gravity coupling effects},}\ }\href
  {\doibase 10.1103/PhysRevLett.113.023005} {\bibfield  {journal} {\bibinfo
  {journal} {Phys. Rev. Lett.}\ }\textbf {\bibinfo {volume} {113}},\ \bibinfo
  {pages} {023005} (\bibinfo {year} {2014})}\BibitemShut {NoStop}%
\bibitem [{\citenamefont {Schlippert}\ \emph {et~al.}(2014)\citenamefont
  {Schlippert}, \citenamefont {Hartwig}, \citenamefont {Albers}, \citenamefont
  {Richardson}, \citenamefont {Schubert}, \citenamefont {Roura}, \citenamefont
  {Schleich}, \citenamefont {Ertmer},\ and\ \citenamefont
  {Rasel}}]{Schlippert2014}%
  \BibitemOpen
  \bibfield  {author} {\bibinfo {author} {\bibfnamefont {D.}~\bibnamefont
  {Schlippert}}, \bibinfo {author} {\bibfnamefont {J.}~\bibnamefont {Hartwig}},
  \bibinfo {author} {\bibfnamefont {H.}~\bibnamefont {Albers}}, \bibinfo
  {author} {\bibfnamefont {L.~L.}\ \bibnamefont {Richardson}}, \bibinfo
  {author} {\bibfnamefont {C.}~\bibnamefont {Schubert}}, \bibinfo {author}
  {\bibfnamefont {A.}~\bibnamefont {Roura}}, \bibinfo {author} {\bibfnamefont
  {W.~P.}\ \bibnamefont {Schleich}}, \bibinfo {author} {\bibfnamefont
  {W.}~\bibnamefont {Ertmer}}, \ and\ \bibinfo {author} {\bibfnamefont {E.~M.}\
  \bibnamefont {Rasel}},\ }\bibfield  {title} {\enquote {\bibinfo {title}
  {Quantum test of the universality of free fall},}\ }\href {\doibase
  10.1103/PhysRevLett.112.203002} {\bibfield  {journal} {\bibinfo  {journal}
  {Phys. Rev. Lett.}\ }\textbf {\bibinfo {volume} {112}},\ \bibinfo {pages}
  {203002} (\bibinfo {year} {2014})}\BibitemShut {NoStop}%
\bibitem [{\citenamefont {Kovachy}\ \emph {et~al.}(2015)\citenamefont
  {Kovachy}, \citenamefont {Asenbaum}, \citenamefont {Overstreet},
  \citenamefont {Donnelly}, \citenamefont {Dickerson}, \citenamefont
  {Sugarbaker}, \citenamefont {Hogan},\ and\ \citenamefont
  {Kasevich}}]{Kovachy2015}%
  \BibitemOpen
  \bibfield  {author} {\bibinfo {author} {\bibfnamefont {T.}~\bibnamefont
  {Kovachy}}, \bibinfo {author} {\bibfnamefont {P.}~\bibnamefont {Asenbaum}},
  \bibinfo {author} {\bibfnamefont {C.}~\bibnamefont {Overstreet}}, \bibinfo
  {author} {\bibfnamefont {C.~A.}\ \bibnamefont {Donnelly}}, \bibinfo {author}
  {\bibfnamefont {S.~M.}\ \bibnamefont {Dickerson}}, \bibinfo {author}
  {\bibfnamefont {A.}~\bibnamefont {Sugarbaker}}, \bibinfo {author}
  {\bibfnamefont {J.~M.}\ \bibnamefont {Hogan}}, \ and\ \bibinfo {author}
  {\bibfnamefont {M.~A.}\ \bibnamefont {Kasevich}},\ }\bibfield  {title}
  {\enquote {\bibinfo {title} {Quantum superposition at the half-metre
  scale},}\ }\href {\doibase 10.1038/nature16155} {\bibfield  {journal}
  {\bibinfo  {journal} {Nature}\ }\textbf {\bibinfo {volume} {528}},\ \bibinfo
  {pages} {530} (\bibinfo {year} {2015})}\BibitemShut {NoStop}%
\bibitem [{\citenamefont {Jaffe}\ \emph {et~al.}(2017)\citenamefont {Jaffe},
  \citenamefont {Haslinger}, \citenamefont {Xu}, \citenamefont {Hamilton},
  \citenamefont {Upadhye}, \citenamefont {Elder}, \citenamefont {Khoury},\ and\
  \citenamefont {M\"{u}ller}}]{Jaffe2017}%
  \BibitemOpen
  \bibfield  {author} {\bibinfo {author} {\bibfnamefont {M.}~\bibnamefont
  {Jaffe}}, \bibinfo {author} {\bibfnamefont {P.}~\bibnamefont {Haslinger}},
  \bibinfo {author} {\bibfnamefont {V.}~\bibnamefont {Xu}}, \bibinfo {author}
  {\bibfnamefont {P.}~\bibnamefont {Hamilton}}, \bibinfo {author}
  {\bibfnamefont {A.}~\bibnamefont {Upadhye}}, \bibinfo {author} {\bibfnamefont
  {B.}~\bibnamefont {Elder}}, \bibinfo {author} {\bibfnamefont
  {J.}~\bibnamefont {Khoury}}, \ and\ \bibinfo {author} {\bibfnamefont
  {H.}~\bibnamefont {M\"{u}ller}},\ }\bibfield  {title} {\enquote {\bibinfo
  {title} {Testing sub-gravitational forces on atoms from a miniature in-vacuum
  source mass},}\ }\href {\doibase 10.1038/nphys4189} {\bibfield  {journal}
  {\bibinfo  {journal} {Nat. Phys.}\ }\textbf {\bibinfo {volume} {13}},\
  \bibinfo {pages} {938} (\bibinfo {year} {2017})}\BibitemShut {NoStop}%
\bibitem [{\citenamefont {Rosi}\ \emph {et~al.}(2017)\citenamefont {Rosi},
  \citenamefont {D’Amico}, \citenamefont {Cacciapuoti}, \citenamefont
  {Sorrentino}, \citenamefont {Prevedelli}, \citenamefont {Zych}, \citenamefont
  {Brukner},\ and\ \citenamefont {Tino}}]{Rosi2017}%
  \BibitemOpen
  \bibfield  {author} {\bibinfo {author} {\bibfnamefont {G.}~\bibnamefont
  {Rosi}}, \bibinfo {author} {\bibfnamefont {G.}~\bibnamefont {D’Amico}},
  \bibinfo {author} {\bibfnamefont {L.}~\bibnamefont {Cacciapuoti}}, \bibinfo
  {author} {\bibfnamefont {F.}~\bibnamefont {Sorrentino}}, \bibinfo {author}
  {\bibfnamefont {M.}~\bibnamefont {Prevedelli}}, \bibinfo {author}
  {\bibfnamefont {M.}~\bibnamefont {Zych}}, \bibinfo {author} {\bibfnamefont
  {\v{C}.}\ \bibnamefont {Brukner}}, \ and\ \bibinfo {author} {\bibfnamefont
  {G.~M.}\ \bibnamefont {Tino}},\ }\bibfield  {title} {\enquote {\bibinfo
  {title} {Quantum test of the equivalence principle for atoms in coherent
  superposition of internal energy states},}\ }\href {\doibase
  10.1038/ncomms15529} {\bibfield  {journal} {\bibinfo  {journal} {Nat. Comm.}\
  }\textbf {\bibinfo {volume} {8}},\ \bibinfo {pages} {15529} (\bibinfo {year}
  {2017})}\BibitemShut {NoStop}%
\bibitem [{\citenamefont {Fixler}\ \emph {et~al.}(2007)\citenamefont {Fixler},
  \citenamefont {Foster}, \citenamefont {McGuirk},\ and\ \citenamefont
  {Kasevich}}]{Fixler74}%
  \BibitemOpen
  \bibfield  {author} {\bibinfo {author} {\bibfnamefont {J.~B.}\ \bibnamefont
  {Fixler}}, \bibinfo {author} {\bibfnamefont {G.~T.}\ \bibnamefont {Foster}},
  \bibinfo {author} {\bibfnamefont {J.~M.}\ \bibnamefont {McGuirk}}, \ and\
  \bibinfo {author} {\bibfnamefont {M.~A.}\ \bibnamefont {Kasevich}},\
  }\bibfield  {title} {\enquote {\bibinfo {title} {Atom interferometer
  measurement of the {N}ewtonian constant of gravity},}\ }\href {\doibase
  10.1126/science.1135459} {\bibfield  {journal} {\bibinfo  {journal}
  {Science}\ }\textbf {\bibinfo {volume} {315}},\ \bibinfo {pages} {74--77}
  (\bibinfo {year} {2007})}\BibitemShut {NoStop}%
\bibitem [{\citenamefont {Hanneke}\ \emph {et~al.}(2008)\citenamefont
  {Hanneke}, \citenamefont {Fogwell},\ and\ \citenamefont
  {Gabrielse}}]{hanneke2008}%
  \BibitemOpen
  \bibfield  {author} {\bibinfo {author} {\bibfnamefont {D.}~\bibnamefont
  {Hanneke}}, \bibinfo {author} {\bibfnamefont {S.}~\bibnamefont {Fogwell}}, \
  and\ \bibinfo {author} {\bibfnamefont {G.}~\bibnamefont {Gabrielse}},\
  }\bibfield  {title} {\enquote {\bibinfo {title} {New measurement of the
  electron magnetic moment and the fine structure constant},}\ }\href {\doibase
  10.1103/PhysRevLett.100.120801} {\bibfield  {journal} {\bibinfo  {journal}
  {Phys. Rev. Lett.}\ }\textbf {\bibinfo {volume} {100}},\ \bibinfo {pages}
  {120801} (\bibinfo {year} {2008})}\BibitemShut {NoStop}%
\bibitem [{\citenamefont {Parker}\ \emph {et~al.}(2018)\citenamefont {Parker},
  \citenamefont {Yu}, \citenamefont {Zhong}, \citenamefont {Estey},\ and\
  \citenamefont {M{\"u}ller}}]{Parker191}%
  \BibitemOpen
  \bibfield  {author} {\bibinfo {author} {\bibfnamefont {R.~H.}\ \bibnamefont
  {Parker}}, \bibinfo {author} {\bibfnamefont {C.}~\bibnamefont {Yu}}, \bibinfo
  {author} {\bibfnamefont {W.}~\bibnamefont {Zhong}}, \bibinfo {author}
  {\bibfnamefont {B.}~\bibnamefont {Estey}}, \ and\ \bibinfo {author}
  {\bibfnamefont {H.}~\bibnamefont {M{\"u}ller}},\ }\bibfield  {title}
  {\enquote {\bibinfo {title} {Measurement of the fine-structure constant as a
  test of the {S}tandard {M}odel},}\ }\href {\doibase 10.1126/science.aap7706}
  {\bibfield  {journal} {\bibinfo  {journal} {Science}\ }\textbf {\bibinfo
  {volume} {360}},\ \bibinfo {pages} {191--195} (\bibinfo {year}
  {2018})}\BibitemShut {NoStop}%
\bibitem [{\citenamefont {Morel}\ \emph {et~al.}(2020)\citenamefont {Morel},
  \citenamefont {Yao}, \citenamefont {Clad\'{e}},\ and\ \citenamefont
  {Guellati-Kh\'{e}lifa}}]{morel2020}%
  \BibitemOpen
  \bibfield  {author} {\bibinfo {author} {\bibfnamefont {L.}~\bibnamefont
  {Morel}}, \bibinfo {author} {\bibfnamefont {Z.}~\bibnamefont {Yao}}, \bibinfo
  {author} {\bibfnamefont {P.}~\bibnamefont {Clad\'{e}}}, \ and\ \bibinfo
  {author} {\bibfnamefont {S.}~\bibnamefont {Guellati-Kh\'{e}lifa}},\
  }\bibfield  {title} {\enquote {\bibinfo {title} {Determination of the
  fine-structure constant with an accuracy of 81 parts per trillion},}\ }\href
  {\doibase 10.1038/s41586-020-2964-7} {\bibfield  {journal} {\bibinfo
  {journal} {Nature}\ }\textbf {\bibinfo {volume} {588}},\ \bibinfo {pages}
  {61--65} (\bibinfo {year} {2020})}\BibitemShut {NoStop}%
\bibitem [{\citenamefont {Xu}\ \emph {et~al.}(2019)\citenamefont {Xu},
  \citenamefont {Jaffe}, \citenamefont {Panda}, \citenamefont {Kristensen},
  \citenamefont {Clark},\ and\ \citenamefont {M{\"u}ller}}]{Xu745}%
  \BibitemOpen
  \bibfield  {author} {\bibinfo {author} {\bibfnamefont {V.}~\bibnamefont
  {Xu}}, \bibinfo {author} {\bibfnamefont {M.}~\bibnamefont {Jaffe}}, \bibinfo
  {author} {\bibfnamefont {C.~D.}\ \bibnamefont {Panda}}, \bibinfo {author}
  {\bibfnamefont {S.~L.}\ \bibnamefont {Kristensen}}, \bibinfo {author}
  {\bibfnamefont {L.~W.}\ \bibnamefont {Clark}}, \ and\ \bibinfo {author}
  {\bibfnamefont {H.}~\bibnamefont {M{\"u}ller}},\ }\bibfield  {title}
  {\enquote {\bibinfo {title} {Probing gravity by holding atoms for 20
  seconds},}\ }\href {\doibase 10.1126/science.aay6428} {\bibfield  {journal}
  {\bibinfo  {journal} {Science}\ }\textbf {\bibinfo {volume} {366}},\ \bibinfo
  {pages} {745--749} (\bibinfo {year} {2019})}\BibitemShut {NoStop}%
\bibitem [{\citenamefont {Barrett}\ \emph {et~al.}(2019)\citenamefont
  {Barrett}, \citenamefont {Cheiney}, \citenamefont {Battelier}, \citenamefont
  {Napolitano},\ and\ \citenamefont {Bouyer}}]{barrett}%
  \BibitemOpen
  \bibfield  {author} {\bibinfo {author} {\bibfnamefont {B.}~\bibnamefont
  {Barrett}}, \bibinfo {author} {\bibfnamefont {P.}~\bibnamefont {Cheiney}},
  \bibinfo {author} {\bibfnamefont {B.}~\bibnamefont {Battelier}}, \bibinfo
  {author} {\bibfnamefont {F.}~\bibnamefont {Napolitano}}, \ and\ \bibinfo
  {author} {\bibfnamefont {P.}~\bibnamefont {Bouyer}},\ }\bibfield  {title}
  {\enquote {\bibinfo {title} {Multidimensional atom optics and
  interferometry},}\ }\href {\doibase 10.1103/PhysRevLett.122.043604}
  {\bibfield  {journal} {\bibinfo  {journal} {Phys. Rev. Lett.}\ }\textbf
  {\bibinfo {volume} {122}},\ \bibinfo {pages} {043604} (\bibinfo {year}
  {2019})}\BibitemShut {NoStop}%
\bibitem [{\citenamefont {M\'{e}noret}\ \emph {et~al.}(2018)\citenamefont
  {M\'{e}noret}, \citenamefont {Vermeulen}, \citenamefont {Le~Moigne},
  \citenamefont {Bonvalot}, \citenamefont {Bouyer}, \citenamefont {Landragin},\
  and\ \citenamefont {Desruelle}}]{Menoret2018}%
  \BibitemOpen
  \bibfield  {author} {\bibinfo {author} {\bibfnamefont {V.}~\bibnamefont
  {M\'{e}noret}}, \bibinfo {author} {\bibfnamefont {P.}~\bibnamefont
  {Vermeulen}}, \bibinfo {author} {\bibfnamefont {N.}~\bibnamefont
  {Le~Moigne}}, \bibinfo {author} {\bibfnamefont {S.}~\bibnamefont {Bonvalot}},
  \bibinfo {author} {\bibfnamefont {P.}~\bibnamefont {Bouyer}}, \bibinfo
  {author} {\bibfnamefont {A.}~\bibnamefont {Landragin}}, \ and\ \bibinfo
  {author} {\bibfnamefont {B.}~\bibnamefont {Desruelle}},\ }\bibfield  {title}
  {\enquote {\bibinfo {title} {Gravity measurements below 10$^{-9}$g with a
  transportable absolute quantum gravimeter},}\ }\href {\doibase
  10.1038/s41598-018-30608-1} {\bibfield  {journal} {\bibinfo  {journal} {Sci.
  Rep.}\ }\textbf {\bibinfo {volume} {8}},\ \bibinfo {pages} {12300} (\bibinfo
  {year} {2018})}\BibitemShut {NoStop}%
\bibitem [{\citenamefont {Bongs}\ \emph {et~al.}(2019)\citenamefont {Bongs},
  \citenamefont {Holynski}, \citenamefont {Vovrosh}, \citenamefont {Bouyer},
  \citenamefont {Condon}, \citenamefont {Rasel}, \citenamefont {Schubert},
  \citenamefont {Schleich},\ and\ \citenamefont {Roura}}]{Bongs2019}%
  \BibitemOpen
  \bibfield  {author} {\bibinfo {author} {\bibfnamefont {K.}~\bibnamefont
  {Bongs}}, \bibinfo {author} {\bibfnamefont {M.}~\bibnamefont {Holynski}},
  \bibinfo {author} {\bibfnamefont {J.}~\bibnamefont {Vovrosh}}, \bibinfo
  {author} {\bibfnamefont {P.}~\bibnamefont {Bouyer}}, \bibinfo {author}
  {\bibfnamefont {G.}~\bibnamefont {Condon}}, \bibinfo {author} {\bibfnamefont
  {E.}~\bibnamefont {Rasel}}, \bibinfo {author} {\bibfnamefont
  {C.}~\bibnamefont {Schubert}}, \bibinfo {author} {\bibfnamefont {W.~P.}\
  \bibnamefont {Schleich}}, \ and\ \bibinfo {author} {\bibfnamefont
  {A.}~\bibnamefont {Roura}},\ }\bibfield  {title} {\enquote {\bibinfo {title}
  {Taking atom interferometric quantum sensors from the laboratory to
  real-world applications},}\ }\href {\doibase 10.1038/s42254-019-0117-4}
  {\bibfield  {journal} {\bibinfo  {journal} {Nat. Rev. Phys.}\ }\textbf
  {\bibinfo {volume} {1}},\ \bibinfo {pages} {731} (\bibinfo {year}
  {2019})}\BibitemShut {NoStop}%
\bibitem [{\citenamefont {Garrido~Alzar}(2019)}]{alzar}%
  \BibitemOpen
  \bibfield  {author} {\bibinfo {author} {\bibfnamefont {C.~L.}\ \bibnamefont
  {Garrido~Alzar}},\ }\bibfield  {title} {\enquote {\bibinfo {title} {Compact
  chip-scale guided cold atom gyrometers for inertial navigation: Enabling
  technologies and design study},}\ }\href {\doibase 10.1116/1.5120348}
  {\bibfield  {journal} {\bibinfo  {journal} {AVS Quantum Science}\ }\textbf
  {\bibinfo {volume} {1}},\ \bibinfo {pages} {014702} (\bibinfo {year}
  {2019})}\BibitemShut {NoStop}%
\bibitem [{\citenamefont {Asenbaum}\ \emph {et~al.}(2017)\citenamefont
  {Asenbaum}, \citenamefont {Overstreet}, \citenamefont {Kovachy},
  \citenamefont {Brown}, \citenamefont {Hogan},\ and\ \citenamefont
  {Kasevich}}]{asenbaum}%
  \BibitemOpen
  \bibfield  {author} {\bibinfo {author} {\bibfnamefont {P.}~\bibnamefont
  {Asenbaum}}, \bibinfo {author} {\bibfnamefont {C.}~\bibnamefont
  {Overstreet}}, \bibinfo {author} {\bibfnamefont {T.}~\bibnamefont {Kovachy}},
  \bibinfo {author} {\bibfnamefont {D.~D.}\ \bibnamefont {Brown}}, \bibinfo
  {author} {\bibfnamefont {J.~M.}\ \bibnamefont {Hogan}}, \ and\ \bibinfo
  {author} {\bibfnamefont {M.~A.}\ \bibnamefont {Kasevich}},\ }\bibfield
  {title} {\enquote {\bibinfo {title} {Phase shift in an atom interferometer
  due to spacetime curvature across its wave function},}\ }\href {\doibase
  10.1103/PhysRevLett.118.183602} {\bibfield  {journal} {\bibinfo  {journal}
  {Phys. Rev. Lett.}\ }\textbf {\bibinfo {volume} {118}},\ \bibinfo {pages}
  {183602} (\bibinfo {year} {2017})}\BibitemShut {NoStop}%
\bibitem [{\citenamefont {Rosi}\ \emph {et~al.}(2014)\citenamefont {Rosi},
  \citenamefont {Sorrentino}, \citenamefont {Cacciapuoti}, \citenamefont
  {Prevedelli},\ and\ \citenamefont {Tino}}]{Rosi2014}%
  \BibitemOpen
  \bibfield  {author} {\bibinfo {author} {\bibfnamefont {G.}~\bibnamefont
  {Rosi}}, \bibinfo {author} {\bibfnamefont {F.}~\bibnamefont {Sorrentino}},
  \bibinfo {author} {\bibfnamefont {L.}~\bibnamefont {Cacciapuoti}}, \bibinfo
  {author} {\bibfnamefont {M.}~\bibnamefont {Prevedelli}}, \ and\ \bibinfo
  {author} {\bibfnamefont {G.~M.}\ \bibnamefont {Tino}},\ }\bibfield  {title}
  {\enquote {\bibinfo {title} {Precision measurement of the {N}ewtonian
  gravitational constant using cold atoms},}\ }\href {\doibase
  10.1038/nature13433} {\bibfield  {journal} {\bibinfo  {journal} {Nature}\
  }\textbf {\bibinfo {volume} {510}},\ \bibinfo {pages} {518} (\bibinfo {year}
  {2014})}\BibitemShut {NoStop}%
\bibitem [{\citenamefont {Hamilton}\ \emph {et~al.}(2015)\citenamefont
  {Hamilton}, \citenamefont {Jaffe}, \citenamefont {Brown}, \citenamefont
  {Maisenbacher}, \citenamefont {Estey},\ and\ \citenamefont
  {M\"uller}}]{hamilton}%
  \BibitemOpen
  \bibfield  {author} {\bibinfo {author} {\bibfnamefont {P.}~\bibnamefont
  {Hamilton}}, \bibinfo {author} {\bibfnamefont {M.}~\bibnamefont {Jaffe}},
  \bibinfo {author} {\bibfnamefont {J.~M.}\ \bibnamefont {Brown}}, \bibinfo
  {author} {\bibfnamefont {L.}~\bibnamefont {Maisenbacher}}, \bibinfo {author}
  {\bibfnamefont {B.}~\bibnamefont {Estey}}, \ and\ \bibinfo {author}
  {\bibfnamefont {H.}~\bibnamefont {M\"uller}},\ }\bibfield  {title} {\enquote
  {\bibinfo {title} {Atom interferometry in an optical cavity},}\ }\href
  {\doibase 10.1103/PhysRevLett.114.100405} {\bibfield  {journal} {\bibinfo
  {journal} {Phys. Rev. Lett.}\ }\textbf {\bibinfo {volume} {114}},\ \bibinfo
  {pages} {100405} (\bibinfo {year} {2015})}\BibitemShut {NoStop}%
\bibitem [{\citenamefont {Dickerson}\ \emph {et~al.}(2013)\citenamefont
  {Dickerson}, \citenamefont {Hogan}, \citenamefont {Sugarbaker}, \citenamefont
  {Johnson},\ and\ \citenamefont {Kasevich}}]{dickerson}%
  \BibitemOpen
  \bibfield  {author} {\bibinfo {author} {\bibfnamefont {S.~M.}\ \bibnamefont
  {Dickerson}}, \bibinfo {author} {\bibfnamefont {J.~M.}\ \bibnamefont
  {Hogan}}, \bibinfo {author} {\bibfnamefont {A.}~\bibnamefont {Sugarbaker}},
  \bibinfo {author} {\bibfnamefont {D.~M.~S.}\ \bibnamefont {Johnson}}, \ and\
  \bibinfo {author} {\bibfnamefont {M.~A.}\ \bibnamefont {Kasevich}},\
  }\bibfield  {title} {\enquote {\bibinfo {title} {Multiaxis inertial sensing
  with long-time point source atom interferometry},}\ }\href {\doibase
  10.1103/PhysRevLett.111.083001} {\bibfield  {journal} {\bibinfo  {journal}
  {Phys. Rev. Lett.}\ }\textbf {\bibinfo {volume} {111}},\ \bibinfo {pages}
  {083001} (\bibinfo {year} {2013})}\BibitemShut {NoStop}%
\bibitem [{\citenamefont {Hoth}\ \emph {et~al.}(2016)\citenamefont {Hoth},
  \citenamefont {Pelle}, \citenamefont {Riedl}, \citenamefont {Kitching},\ and\
  \citenamefont {Donley}}]{hoth}%
  \BibitemOpen
  \bibfield  {author} {\bibinfo {author} {\bibfnamefont {G.~W.}\ \bibnamefont
  {Hoth}}, \bibinfo {author} {\bibfnamefont {B.}~\bibnamefont {Pelle}},
  \bibinfo {author} {\bibfnamefont {S.}~\bibnamefont {Riedl}}, \bibinfo
  {author} {\bibfnamefont {J.}~\bibnamefont {Kitching}}, \ and\ \bibinfo
  {author} {\bibfnamefont {E.~A.}\ \bibnamefont {Donley}},\ }\bibfield  {title}
  {\enquote {\bibinfo {title} {Point source atom interferometry with a cloud of
  finite size},}\ }\href {\doibase 10.1063/1.4961527} {\bibfield  {journal}
  {\bibinfo  {journal} {Appl. Phys. Lett.}\ }\textbf {\bibinfo {volume}
  {109}},\ \bibinfo {pages} {071113} (\bibinfo {year} {2016})}\BibitemShut
  {NoStop}%
\bibitem [{\citenamefont {Wu}\ \emph {et~al.}(2007)\citenamefont {Wu},
  \citenamefont {Su},\ and\ \citenamefont {Prentiss}}]{wu2007}%
  \BibitemOpen
  \bibfield  {author} {\bibinfo {author} {\bibfnamefont {S.}~\bibnamefont
  {Wu}}, \bibinfo {author} {\bibfnamefont {E.}~\bibnamefont {Su}}, \ and\
  \bibinfo {author} {\bibfnamefont {M.}~\bibnamefont {Prentiss}},\ }\bibfield
  {title} {\enquote {\bibinfo {title} {Demonstration of an area-enclosing
  guided-atom interferometer for rotation sensing},}\ }\href {\doibase
  10.1103/PhysRevLett.99.173201} {\bibfield  {journal} {\bibinfo  {journal}
  {Phys. Rev. Lett.}\ }\textbf {\bibinfo {volume} {99}},\ \bibinfo {pages}
  {173201} (\bibinfo {year} {2007})}\BibitemShut {NoStop}%
\bibitem [{\citenamefont {Moan}\ \emph {et~al.}(2020)\citenamefont {Moan},
  \citenamefont {Horne}, \citenamefont {Arpornthip}, \citenamefont {Luo},
  \citenamefont {Fallon}, \citenamefont {Berl},\ and\ \citenamefont
  {Sackett}}]{moan2020}%
  \BibitemOpen
  \bibfield  {author} {\bibinfo {author} {\bibfnamefont {E.~R.}\ \bibnamefont
  {Moan}}, \bibinfo {author} {\bibfnamefont {R.~A.}\ \bibnamefont {Horne}},
  \bibinfo {author} {\bibfnamefont {T.}~\bibnamefont {Arpornthip}}, \bibinfo
  {author} {\bibfnamefont {Z.}~\bibnamefont {Luo}}, \bibinfo {author}
  {\bibfnamefont {A.~J.}\ \bibnamefont {Fallon}}, \bibinfo {author}
  {\bibfnamefont {S.~J.}\ \bibnamefont {Berl}}, \ and\ \bibinfo {author}
  {\bibfnamefont {C.~A.}\ \bibnamefont {Sackett}},\ }\bibfield  {title}
  {\enquote {\bibinfo {title} {Quantum rotation sensing with dual sagnac
  interferometers in an atom-optical waveguide},}\ }\href {\doibase
  10.1103/PhysRevLett.124.120403} {\bibfield  {journal} {\bibinfo  {journal}
  {Phys. Rev. Lett.}\ }\textbf {\bibinfo {volume} {124}},\ \bibinfo {pages}
  {120403} (\bibinfo {year} {2020})}\BibitemShut {NoStop}%
\bibitem [{\citenamefont {Davis}\ and\ \citenamefont
  {Narducci}(2008)}]{Davis2008}%
  \BibitemOpen
  \bibfield  {author} {\bibinfo {author} {\bibfnamefont {J.P.}\ \bibnamefont
  {Davis}}\ and\ \bibinfo {author} {\bibfnamefont {F.A.}\ \bibnamefont
  {Narducci}},\ }\bibfield  {title} {\enquote {\bibinfo {title} {A proposal for
  a gradient magnetometer atom interferometer},}\ }\href
  {http://dx.doi.org/10.1080/09500340802468633} {\bibfield  {journal} {\bibinfo
   {journal} {J. Mod. Opt.}\ }\textbf {\bibinfo {volume} {55}},\ \bibinfo
  {pages} {3173} (\bibinfo {year} {2008})}\BibitemShut {NoStop}%
\bibitem [{\citenamefont {Zimmermann}\ \emph {et~al.}(2019)\citenamefont
  {Zimmermann}, \citenamefont {Efremov}, \citenamefont {Zeller}, \citenamefont
  {Schleich}, \citenamefont {Davis},\ and\ \citenamefont
  {Narducci}}]{Zimmermann2019}%
  \BibitemOpen
  \bibfield  {author} {\bibinfo {author} {\bibfnamefont {M.}~\bibnamefont
  {Zimmermann}}, \bibinfo {author} {\bibfnamefont {M.A.}\ \bibnamefont
  {Efremov}}, \bibinfo {author} {\bibfnamefont {W.}~\bibnamefont {Zeller}},
  \bibinfo {author} {\bibfnamefont {W.P.}\ \bibnamefont {Schleich}}, \bibinfo
  {author} {\bibfnamefont {J.P.}\ \bibnamefont {Davis}}, \ and\ \bibinfo
  {author} {\bibfnamefont {F.A.}\ \bibnamefont {Narducci}},\ }\bibfield
  {title} {\enquote {\bibinfo {title} {Representation-free description of atom
  interferometers in time-dependent linear potentials},}\ }\href
  {http://dx.doi.org/10.1088/1367-2630/ab2e8c} {\bibfield  {journal} {\bibinfo
  {journal} {New J. Phys.}\ }\textbf {\bibinfo {volume} {21}},\ \bibinfo
  {pages} {073031} (\bibinfo {year} {2019})}\BibitemShut {NoStop}%
\bibitem [{\citenamefont {Chen}\ \emph {et~al.}(2020)\citenamefont {Chen},
  \citenamefont {Hansen}, \citenamefont {Shuker}, \citenamefont {Boudot},
  \citenamefont {Kitching},\ and\ \citenamefont {Donley}}]{chen2020}%
  \BibitemOpen
  \bibfield  {author} {\bibinfo {author} {\bibfnamefont {Y.-J.}\ \bibnamefont
  {Chen}}, \bibinfo {author} {\bibfnamefont {A.}~\bibnamefont {Hansen}},
  \bibinfo {author} {\bibfnamefont {M.}~\bibnamefont {Shuker}}, \bibinfo
  {author} {\bibfnamefont {R.}~\bibnamefont {Boudot}}, \bibinfo {author}
  {\bibfnamefont {J.}~\bibnamefont {Kitching}}, \ and\ \bibinfo {author}
  {\bibfnamefont {E.A.}\ \bibnamefont {Donley}},\ }\bibfield  {title} {\enquote
  {\bibinfo {title} {Robust inertial sensing with point-source atom
  interferometry for interferograms spanning a partial period},}\ }\href
  {\doibase 10.1364/OE.399988} {\bibfield  {journal} {\bibinfo  {journal} {Opt.
  Express}\ }\textbf {\bibinfo {volume} {28}},\ \bibinfo {pages} {34516--34529}
  (\bibinfo {year} {2020})}\BibitemShut {NoStop}%
\bibitem [{\citenamefont {Stickney}\ and\ \citenamefont
  {Zozulya}(2002)}]{stickney2002}%
  \BibitemOpen
  \bibfield  {author} {\bibinfo {author} {\bibfnamefont {J.~A.}\ \bibnamefont
  {Stickney}}\ and\ \bibinfo {author} {\bibfnamefont {A.~A.}\ \bibnamefont
  {Zozulya}},\ }\bibfield  {title} {\enquote {\bibinfo {title} {Wave-function
  recombination instability in cold-atom interferometers},}\ }\href {\doibase
  10.1103/PhysRevA.66.053601} {\bibfield  {journal} {\bibinfo  {journal} {Phys.
  Rev. A}\ }\textbf {\bibinfo {volume} {66}},\ \bibinfo {pages} {053601}
  (\bibinfo {year} {2002})}\BibitemShut {NoStop}%
\bibitem [{\citenamefont {Stickney}\ and\ \citenamefont
  {Zozulya}(2003)}]{stickney}%
  \BibitemOpen
  \bibfield  {author} {\bibinfo {author} {\bibfnamefont {J.~A.}\ \bibnamefont
  {Stickney}}\ and\ \bibinfo {author} {\bibfnamefont {A.~A.}\ \bibnamefont
  {Zozulya}},\ }\bibfield  {title} {\enquote {\bibinfo {title} {Influence of
  nonadiabaticity and nonlinearity on the operation of cold-atom beam
  splitters},}\ }\href {\doibase 10.1103/PhysRevA.68.013611} {\bibfield
  {journal} {\bibinfo  {journal} {Phys. Rev. A}\ }\textbf {\bibinfo {volume}
  {68}},\ \bibinfo {pages} {013611} (\bibinfo {year} {2003})}\BibitemShut
  {NoStop}%
\bibitem [{\citenamefont {Burke}\ \emph {et~al.}(2008)\citenamefont {Burke},
  \citenamefont {Deissler}, \citenamefont {Hughes},\ and\ \citenamefont
  {Sackett}}]{burke}%
  \BibitemOpen
  \bibfield  {author} {\bibinfo {author} {\bibfnamefont {J.~H.~T.}\
  \bibnamefont {Burke}}, \bibinfo {author} {\bibfnamefont {B.}~\bibnamefont
  {Deissler}}, \bibinfo {author} {\bibfnamefont {K.~J.}\ \bibnamefont
  {Hughes}}, \ and\ \bibinfo {author} {\bibfnamefont {C.~A.}\ \bibnamefont
  {Sackett}},\ }\bibfield  {title} {\enquote {\bibinfo {title} {Confinement
  effects in a guided-wave atom interferometer with millimeter-scale arm
  separation},}\ }\href {\doibase 10.1103/PhysRevA.78.023619} {\bibfield
  {journal} {\bibinfo  {journal} {Phys. Rev. A}\ }\textbf {\bibinfo {volume}
  {78}},\ \bibinfo {pages} {023619} (\bibinfo {year} {2008})}\BibitemShut
  {NoStop}%
\bibitem [{\citenamefont {Stickney}\ \emph {et~al.}(2008)\citenamefont
  {Stickney}, \citenamefont {Kafle}, \citenamefont {Anderson},\ and\
  \citenamefont {Zozulya}}]{stickney2008}%
  \BibitemOpen
  \bibfield  {author} {\bibinfo {author} {\bibfnamefont {J.~A.}\ \bibnamefont
  {Stickney}}, \bibinfo {author} {\bibfnamefont {R.~P.}\ \bibnamefont {Kafle}},
  \bibinfo {author} {\bibfnamefont {D.~Z.}\ \bibnamefont {Anderson}}, \ and\
  \bibinfo {author} {\bibfnamefont {A.~A.}\ \bibnamefont {Zozulya}},\
  }\bibfield  {title} {\enquote {\bibinfo {title} {Theoretical analysis of a
  single- and double-reflection atom interferometer in a weakly confining
  magnetic trap},}\ }\href {\doibase 10.1103/PhysRevA.77.043604} {\bibfield
  {journal} {\bibinfo  {journal} {Phys. Rev. A}\ }\textbf {\bibinfo {volume}
  {77}},\ \bibinfo {pages} {043604} (\bibinfo {year} {2008})}\BibitemShut
  {NoStop}%
\bibitem [{\citenamefont {Koonin}\ and\ \citenamefont
  {Meredith}(1990)}]{cnaref}%
  \BibitemOpen
  \bibfield  {author} {\bibinfo {author} {\bibfnamefont {S.E.}\ \bibnamefont
  {Koonin}}\ and\ \bibinfo {author} {\bibfnamefont {D.C.}\ \bibnamefont
  {Meredith}},\ }\enquote {\bibinfo {title} {Computational physics {F}ortran
  version},}\ \ (\bibinfo  {publisher} {Addison-Wesley},\ \bibinfo {address}
  {Reading, MA},\ \bibinfo {year} {1990})\ pp.\ \bibinfo {pages}
  {169--180}\BibitemShut {NoStop}%
\bibitem [{\citenamefont {Shin}\ \emph {et~al.}(2004)\citenamefont {Shin},
  \citenamefont {Saba}, \citenamefont {Pasquini}, \citenamefont {Ketterle},
  \citenamefont {Pritchard},\ and\ \citenamefont {Leanhardt}}]{shin}%
  \BibitemOpen
  \bibfield  {author} {\bibinfo {author} {\bibfnamefont {Y.}~\bibnamefont
  {Shin}}, \bibinfo {author} {\bibfnamefont {M.}~\bibnamefont {Saba}}, \bibinfo
  {author} {\bibfnamefont {T.~A.}\ \bibnamefont {Pasquini}}, \bibinfo {author}
  {\bibfnamefont {W.}~\bibnamefont {Ketterle}}, \bibinfo {author}
  {\bibfnamefont {D.~E.}\ \bibnamefont {Pritchard}}, \ and\ \bibinfo {author}
  {\bibfnamefont {A.~E.}\ \bibnamefont {Leanhardt}},\ }\bibfield  {title}
  {\enquote {\bibinfo {title} {Atom interferometry with {B}ose-{E}instein
  condensates in a double-well potential},}\ }\href {\doibase
  10.1103/PhysRevLett.92.050405} {\bibfield  {journal} {\bibinfo  {journal}
  {Phys. Rev. Lett.}\ }\textbf {\bibinfo {volume} {92}},\ \bibinfo {pages}
  {050405} (\bibinfo {year} {2004})}\BibitemShut {NoStop}%
\bibitem [{\citenamefont {H\"ansel}\ \emph
  {et~al.}(2001{\natexlab{a}})\citenamefont {H\"ansel}, \citenamefont
  {Reichel}, \citenamefont {Hommelhoff},\ and\ \citenamefont
  {H\"ansch}}]{hansel}%
  \BibitemOpen
  \bibfield  {author} {\bibinfo {author} {\bibfnamefont {W.}~\bibnamefont
  {H\"ansel}}, \bibinfo {author} {\bibfnamefont {J.}~\bibnamefont {Reichel}},
  \bibinfo {author} {\bibfnamefont {P.}~\bibnamefont {Hommelhoff}}, \ and\
  \bibinfo {author} {\bibfnamefont {T.~W.}\ \bibnamefont {H\"ansch}},\
  }\bibfield  {title} {\enquote {\bibinfo {title} {Magnetic conveyor belt for
  transporting and merging trapped atom clouds},}\ }\href {\doibase
  10.1103/PhysRevLett.86.608} {\bibfield  {journal} {\bibinfo  {journal} {Phys.
  Rev. Lett.}\ }\textbf {\bibinfo {volume} {86}},\ \bibinfo {pages} {608--611}
  (\bibinfo {year} {2001}{\natexlab{a}})}\BibitemShut {NoStop}%
\bibitem [{\citenamefont {Hinds}\ \emph {et~al.}(2001)\citenamefont {Hinds},
  \citenamefont {Vale},\ and\ \citenamefont {Boshier}}]{hinds}%
  \BibitemOpen
  \bibfield  {author} {\bibinfo {author} {\bibfnamefont {E.~A.}\ \bibnamefont
  {Hinds}}, \bibinfo {author} {\bibfnamefont {C.~J.}\ \bibnamefont {Vale}}, \
  and\ \bibinfo {author} {\bibfnamefont {M.~G.}\ \bibnamefont {Boshier}},\
  }\bibfield  {title} {\enquote {\bibinfo {title} {Two-wire waveguide and
  interferometer for cold atoms},}\ }\href {\doibase
  10.1103/PhysRevLett.86.1462} {\bibfield  {journal} {\bibinfo  {journal}
  {Phys. Rev. Lett.}\ }\textbf {\bibinfo {volume} {86}},\ \bibinfo {pages}
  {1462--1465} (\bibinfo {year} {2001})}\BibitemShut {NoStop}%
\bibitem [{\citenamefont {H\"ansel}\ \emph
  {et~al.}(2001{\natexlab{b}})\citenamefont {H\"ansel}, \citenamefont
  {Reichel}, \citenamefont {Hommelhoff},\ and\ \citenamefont
  {H\"ansch}}]{hanselpra}%
  \BibitemOpen
  \bibfield  {author} {\bibinfo {author} {\bibfnamefont {W.}~\bibnamefont
  {H\"ansel}}, \bibinfo {author} {\bibfnamefont {J.}~\bibnamefont {Reichel}},
  \bibinfo {author} {\bibfnamefont {P.}~\bibnamefont {Hommelhoff}}, \ and\
  \bibinfo {author} {\bibfnamefont {T.~W.}\ \bibnamefont {H\"ansch}},\
  }\bibfield  {title} {\enquote {\bibinfo {title} {Trapped-atom interferometer
  in a magnetic microtrap},}\ }\href {\doibase 10.1103/PhysRevA.64.063607}
  {\bibfield  {journal} {\bibinfo  {journal} {Phys. Rev. A}\ }\textbf {\bibinfo
  {volume} {64}},\ \bibinfo {pages} {063607} (\bibinfo {year}
  {2001}{\natexlab{b}})}\BibitemShut {NoStop}%
\bibitem [{\citenamefont {Endres}\ \emph {et~al.}(2016)\citenamefont {Endres},
  \citenamefont {Bernien}, \citenamefont {Keesling}, \citenamefont {Levine},
  \citenamefont {Anschuetz}, \citenamefont {Krajenbrink}, \citenamefont
  {Senko}, \citenamefont {Vuletic}, \citenamefont {Greiner},\ and\
  \citenamefont {Lukin}}]{Endres1024}%
  \BibitemOpen
  \bibfield  {author} {\bibinfo {author} {\bibfnamefont {M.}~\bibnamefont
  {Endres}}, \bibinfo {author} {\bibfnamefont {H.}~\bibnamefont {Bernien}},
  \bibinfo {author} {\bibfnamefont {A.}~\bibnamefont {Keesling}}, \bibinfo
  {author} {\bibfnamefont {H.}~\bibnamefont {Levine}}, \bibinfo {author}
  {\bibfnamefont {E.~R.}\ \bibnamefont {Anschuetz}}, \bibinfo {author}
  {\bibfnamefont {A.}~\bibnamefont {Krajenbrink}}, \bibinfo {author}
  {\bibfnamefont {C.}~\bibnamefont {Senko}}, \bibinfo {author} {\bibfnamefont
  {V.}~\bibnamefont {Vuletic}}, \bibinfo {author} {\bibfnamefont
  {M.}~\bibnamefont {Greiner}}, \ and\ \bibinfo {author} {\bibfnamefont
  {M.~D.}\ \bibnamefont {Lukin}},\ }\bibfield  {title} {\enquote {\bibinfo
  {title} {Atom-by-atom assembly of defect-free one-dimensional cold atom
  arrays},}\ }\href {\doibase 10.1126/science.aah3752} {\bibfield  {journal}
  {\bibinfo  {journal} {Science}\ }\textbf {\bibinfo {volume} {354}},\ \bibinfo
  {pages} {1024--1027} (\bibinfo {year} {2016})}\BibitemShut {NoStop}%
\bibitem [{\citenamefont {Barredo}\ \emph {et~al.}(2016)\citenamefont
  {Barredo}, \citenamefont {de~L{\'e}s{\'e}leuc}, \citenamefont {Lienhard},
  \citenamefont {Lahaye},\ and\ \citenamefont {Browaeys}}]{Barredo1021}%
  \BibitemOpen
  \bibfield  {author} {\bibinfo {author} {\bibfnamefont {D.}~\bibnamefont
  {Barredo}}, \bibinfo {author} {\bibfnamefont {S.}~\bibnamefont
  {de~L{\'e}s{\'e}leuc}}, \bibinfo {author} {\bibfnamefont {V.}~\bibnamefont
  {Lienhard}}, \bibinfo {author} {\bibfnamefont {T.}~\bibnamefont {Lahaye}}, \
  and\ \bibinfo {author} {\bibfnamefont {A.}~\bibnamefont {Browaeys}},\
  }\bibfield  {title} {\enquote {\bibinfo {title} {An atom-by-atom assembler of
  defect-free arbitrary two-dimensional atomic arrays},}\ }\href {\doibase
  10.1126/science.aah3778} {\bibfield  {journal} {\bibinfo  {journal}
  {Science}\ }\textbf {\bibinfo {volume} {354}},\ \bibinfo {pages} {1021--1023}
  (\bibinfo {year} {2016})}\BibitemShut {NoStop}%
\bibitem [{\citenamefont {AU~Kim}\ \emph {et~al.}(2016)\citenamefont {AU~Kim},
  \citenamefont {Lee}, \citenamefont {Lee}, \citenamefont {Jo}, \citenamefont
  {Song},\ and\ \citenamefont {Ahn}}]{kim2016}%
  \BibitemOpen
  \bibfield  {author} {\bibinfo {author} {\bibfnamefont {H.}~\bibnamefont
  {AU~Kim}}, \bibinfo {author} {\bibfnamefont {W.}~\bibnamefont {Lee}},
  \bibinfo {author} {\bibfnamefont {H.}~\bibnamefont {Lee}}, \bibinfo {author}
  {\bibfnamefont {H.}~\bibnamefont {Jo}}, \bibinfo {author} {\bibfnamefont
  {Y.}~\bibnamefont {Song}}, \ and\ \bibinfo {author} {\bibfnamefont
  {J.}~\bibnamefont {Ahn}},\ }\bibfield  {title} {\enquote {\bibinfo {title}
  {In situ single-atom array synthesis using dynamic holographic optical
  tweezers},}\ }\href {\doibase 10.1038/ncomms13317} {\bibfield  {journal}
  {\bibinfo  {journal} {Nat. Comm.}\ }\textbf {\bibinfo {volume} {7}},\
  \bibinfo {pages} {13317} (\bibinfo {year} {2016})}\BibitemShut {NoStop}%
\bibitem [{\citenamefont {Ohl~de Mello}\ \emph {et~al.}(2019)\citenamefont
  {Ohl~de Mello}, \citenamefont {Sch\"affner}, \citenamefont {Werkmann},
  \citenamefont {Preuschoff}, \citenamefont {Kohfahl}, \citenamefont
  {Schlosser},\ and\ \citenamefont {Birkl}}]{ohl}%
  \BibitemOpen
  \bibfield  {author} {\bibinfo {author} {\bibfnamefont {D.}~\bibnamefont
  {Ohl~de Mello}}, \bibinfo {author} {\bibfnamefont {D.}~\bibnamefont
  {Sch\"affner}}, \bibinfo {author} {\bibfnamefont {J.}~\bibnamefont
  {Werkmann}}, \bibinfo {author} {\bibfnamefont {T.}~\bibnamefont
  {Preuschoff}}, \bibinfo {author} {\bibfnamefont {L.}~\bibnamefont {Kohfahl}},
  \bibinfo {author} {\bibfnamefont {M.}~\bibnamefont {Schlosser}}, \ and\
  \bibinfo {author} {\bibfnamefont {G.}~\bibnamefont {Birkl}},\ }\bibfield
  {title} {\enquote {\bibinfo {title} {Defect-free assembly of 2{D} clusters of
  more than 100 single-atom quantum systems},}\ }\href {\doibase
  10.1103/PhysRevLett.122.203601} {\bibfield  {journal} {\bibinfo  {journal}
  {Phys. Rev. Lett.}\ }\textbf {\bibinfo {volume} {122}},\ \bibinfo {pages}
  {203601} (\bibinfo {year} {2019})}\BibitemShut {NoStop}%
\bibitem [{\citenamefont {Schymik}\ \emph {et~al.}(2020)\citenamefont
  {Schymik}, \citenamefont {Lienhard}, \citenamefont {Barredo}, \citenamefont
  {Scholl}, \citenamefont {Williams}, \citenamefont {Browaeys},\ and\
  \citenamefont {Lahaye}}]{schymik}%
  \BibitemOpen
  \bibfield  {author} {\bibinfo {author} {\bibfnamefont {K.-N.}\ \bibnamefont
  {Schymik}}, \bibinfo {author} {\bibfnamefont {V.}~\bibnamefont {Lienhard}},
  \bibinfo {author} {\bibfnamefont {D.}~\bibnamefont {Barredo}}, \bibinfo
  {author} {\bibfnamefont {P.}~\bibnamefont {Scholl}}, \bibinfo {author}
  {\bibfnamefont {H.}~\bibnamefont {Williams}}, \bibinfo {author}
  {\bibfnamefont {A.}~\bibnamefont {Browaeys}}, \ and\ \bibinfo {author}
  {\bibfnamefont {T.}~\bibnamefont {Lahaye}},\ }\bibfield  {title} {\enquote
  {\bibinfo {title} {Enhanced atom-by-atom assembly of arbitrary tweezer
  arrays},}\ }\href {\doibase 10.1103/PhysRevA.102.063107} {\bibfield
  {journal} {\bibinfo  {journal} {Phys. Rev. A}\ }\textbf {\bibinfo {volume}
  {102}},\ \bibinfo {pages} {063107} (\bibinfo {year} {2020})}\BibitemShut
  {NoStop}%
\bibitem [{\citenamefont {Kovachy}\ \emph {et~al.}(2010)\citenamefont
  {Kovachy}, \citenamefont {Hogan}, \citenamefont {Johnson},\ and\
  \citenamefont {Kasevich}}]{kovachypra2010}%
  \BibitemOpen
  \bibfield  {author} {\bibinfo {author} {\bibfnamefont {T.}~\bibnamefont
  {Kovachy}}, \bibinfo {author} {\bibfnamefont {J.~M.}\ \bibnamefont {Hogan}},
  \bibinfo {author} {\bibfnamefont {D.~M.~S.}\ \bibnamefont {Johnson}}, \ and\
  \bibinfo {author} {\bibfnamefont {M.~A.}\ \bibnamefont {Kasevich}},\
  }\bibfield  {title} {\enquote {\bibinfo {title} {Optical lattices as
  waveguides and beam splitters for atom interferometry: An analytical
  treatment and proposal of applications},}\ }\href {\doibase
  10.1103/PhysRevA.82.013638} {\bibfield  {journal} {\bibinfo  {journal} {Phys.
  Rev. A}\ }\textbf {\bibinfo {volume} {82}},\ \bibinfo {pages} {013638}
  (\bibinfo {year} {2010})}\BibitemShut {NoStop}%
\bibitem [{\citenamefont {Mandel}\ \emph {et~al.}(2003)\citenamefont {Mandel},
  \citenamefont {Greiner}, \citenamefont {Widera}, \citenamefont {Rom},
  \citenamefont {H\"ansch},\ and\ \citenamefont {Bloch}}]{Mandel2003}%
  \BibitemOpen
  \bibfield  {author} {\bibinfo {author} {\bibfnamefont {O.}~\bibnamefont
  {Mandel}}, \bibinfo {author} {\bibfnamefont {M.}~\bibnamefont {Greiner}},
  \bibinfo {author} {\bibfnamefont {A.}~\bibnamefont {Widera}}, \bibinfo
  {author} {\bibfnamefont {T.}~\bibnamefont {Rom}}, \bibinfo {author}
  {\bibfnamefont {T.~W.}\ \bibnamefont {H\"ansch}}, \ and\ \bibinfo {author}
  {\bibfnamefont {I.}~\bibnamefont {Bloch}},\ }\bibfield  {title} {\enquote
  {\bibinfo {title} {Coherent transport of neutral atoms in spin-dependent
  optical lattice potentials},}\ }\href {\doibase
  10.1103/PhysRevLett.91.010407} {\bibfield  {journal} {\bibinfo  {journal}
  {Phys. Rev. Lett.}\ }\textbf {\bibinfo {volume} {91}},\ \bibinfo {pages}
  {010407} (\bibinfo {year} {2003})}\BibitemShut {NoStop}%
\bibitem [{\citenamefont {Kuhr}\ \emph {et~al.}(2001)\citenamefont {Kuhr},
  \citenamefont {Alt}, \citenamefont {Schrader}, \citenamefont {M{\"u}ller},
  \citenamefont {Gomer},\ and\ \citenamefont {Meschede}}]{Kuhr2001}%
  \BibitemOpen
  \bibfield  {author} {\bibinfo {author} {\bibfnamefont {S.}~\bibnamefont
  {Kuhr}}, \bibinfo {author} {\bibfnamefont {W.}~\bibnamefont {Alt}}, \bibinfo
  {author} {\bibfnamefont {D.}~\bibnamefont {Schrader}}, \bibinfo {author}
  {\bibfnamefont {M.}~\bibnamefont {M{\"u}ller}}, \bibinfo {author}
  {\bibfnamefont {V.}~\bibnamefont {Gomer}}, \ and\ \bibinfo {author}
  {\bibfnamefont {D.}~\bibnamefont {Meschede}},\ }\bibfield  {title} {\enquote
  {\bibinfo {title} {Deterministic delivery of a single atom},}\ }\href
  {\doibase 10.1126/science.1062725} {\bibfield  {journal} {\bibinfo  {journal}
  {Science}\ }\textbf {\bibinfo {volume} {293}},\ \bibinfo {pages} {278--280}
  (\bibinfo {year} {2001})}\BibitemShut {NoStop}%
\bibitem [{\citenamefont {Schmid}\ \emph {et~al.}(2006)\citenamefont {Schmid},
  \citenamefont {Thalhammer}, \citenamefont {Winkler}, \citenamefont {Lang},\
  and\ \citenamefont {Denschlag}}]{Schmid2006}%
  \BibitemOpen
  \bibfield  {author} {\bibinfo {author} {\bibfnamefont {S.}~\bibnamefont
  {Schmid}}, \bibinfo {author} {\bibfnamefont {G.}~\bibnamefont {Thalhammer}},
  \bibinfo {author} {\bibfnamefont {K.}~\bibnamefont {Winkler}}, \bibinfo
  {author} {\bibfnamefont {F.}~\bibnamefont {Lang}}, \ and\ \bibinfo {author}
  {\bibfnamefont {J.~H.}\ \bibnamefont {Denschlag}},\ }\bibfield  {title}
  {\enquote {\bibinfo {title} {Long distance transport of ultracold atoms using
  a 1{D} optical lattice},}\ }\href {\doibase 10.1088/1367-2630/8/8/159}
  {\bibfield  {journal} {\bibinfo  {journal} {New J. Phys.}\ }\textbf {\bibinfo
  {volume} {8}},\ \bibinfo {pages} {159--159} (\bibinfo {year}
  {2006})}\BibitemShut {NoStop}%
\bibitem [{\citenamefont {Kumar}\ \emph {et~al.}(2018)\citenamefont {Kumar},
  \citenamefont {Wu}, \citenamefont {Giraldo},\ and\ \citenamefont
  {Weiss}}]{Kumar2018}%
  \BibitemOpen
  \bibfield  {author} {\bibinfo {author} {\bibfnamefont {A.}~\bibnamefont
  {Kumar}}, \bibinfo {author} {\bibfnamefont {T.-Y.}\ \bibnamefont {Wu}},
  \bibinfo {author} {\bibfnamefont {F.}~\bibnamefont {Giraldo}}, \ and\
  \bibinfo {author} {\bibfnamefont {D.~S.}\ \bibnamefont {Weiss}},\ }\bibfield
  {title} {\enquote {\bibinfo {title} {Sorting ultracold atoms in a
  three-dimensional optical lattice in a realization of {M}axwell’s demon},}\
  }\href {\doibase 10.1038/s41586-018-0458-7} {\bibfield  {journal} {\bibinfo
  {journal} {Nature}\ }\textbf {\bibinfo {volume} {561}},\ \bibinfo {pages}
  {83} (\bibinfo {year} {2018})}\BibitemShut {NoStop}%
\bibitem [{\citenamefont {Chiow}\ \emph {et~al.}(2009)\citenamefont {Chiow},
  \citenamefont {Herrmann}, \citenamefont {Chu},\ and\ \citenamefont
  {M\"uller}}]{chiow2009}%
  \BibitemOpen
  \bibfield  {author} {\bibinfo {author} {\bibfnamefont {S.~W.}\ \bibnamefont
  {Chiow}}, \bibinfo {author} {\bibfnamefont {S.}~\bibnamefont {Herrmann}},
  \bibinfo {author} {\bibfnamefont {S.}~\bibnamefont {Chu}}, \ and\ \bibinfo
  {author} {\bibfnamefont {H.}~\bibnamefont {M\"uller}},\ }\bibfield  {title}
  {\enquote {\bibinfo {title} {Noise-immune conjugate large-area atom
  interferometers},}\ }\href {\doibase 10.1103/PhysRevLett.103.050402}
  {\bibfield  {journal} {\bibinfo  {journal} {Phys. Rev. Lett.}\ }\textbf
  {\bibinfo {volume} {103}},\ \bibinfo {pages} {050402} (\bibinfo {year}
  {2009})}\BibitemShut {NoStop}%
\end{thebibliography}%
\end{document}